\documentclass{jfm}
\usepackage{graphicx}
\usepackage{epstopdf, epsfig}
\usepackage{amsmath}
\usepackage{mathrsfs} 
\usepackage{minitoc}
\usepackage{tikz, tikz-3dplot}
\usepackage{pgfplots}
\usepackage[utf8]{inputenc}

\usepackage{accents}
\newcommand{\dbtilde}[1]{\accentset{\approx}{#1}}

 \usepackage[T1]{fontenc}
\renewcommand{\vec}[1]{\mbox{\boldmath $ #1 $}}

\renewcommand{\vec}[1]{\mbox{\boldmath $ #1 $}}

\newcommand{\ii}{\mathrm i}
\newcommand{\ee}{\mathrm e}
\newcommand{\intz}{\int_0^{\infty}}
\newcommand{\intt}{\int_{-\infty}^{T_f}}

\shorttitle{Stability of the solitary wave boundary layer}
\shortauthor{A. \"Onder, P. L.-F. Liu}

\title{Stability of the solitary wave boundary layer subject to finite amplitude disturbances}

\author  {Asim \"Onder \aff{1}
  \corresp{\email{asim.onder@gmail.com}},
Philip L.-F. Liu \aff{1,2,3}
  }

\affiliation{ 
\aff{1 }
Department of Civil and Environmental Engineering, National University of Singapore, Singapore 117576, Singapore
\aff{2}
School of Civil and Environmental Engineering, Cornell University, Ithaca, NY 14850, USA
\aff{3}
Institute of Hydrological and Oceanic Sciences, National Central University, Jhongli, Taoyuan, 320, Taiwan}

\begin{document}

\maketitle

\begin{abstract}
The stability and transition in the bottom boundary layer under a solitary wave are analysed in the presence of finite amplitude disturbances. First, the receptivity of the boundary layer is investigated using a linear input-output analysis, in which the environment noise is modelled as distributed body forces. The most dangerous perturbations in a time frame until flow reversal are found to be arranged as counter-rotating streamwise-constant rollers.  One of these roller configurations is then selected and deployed to nonlinear equations, and streaks of various amplitudes are generated via lift-up mechanism. By means of secondary stability analysis and direct numerical simulations, the dual role of streaks in the boundary-layer transition is shown. When the amplitude of streaks remains moderate, these elongated features remain stable until the adverse-pressure-gradient stage and have a dampening effect on the instabilities developing thereafter. In contrast, when the low-speed streaks reach high amplitudes exceeding 15\% of free-stream velocity at the respective phase, they become highly unstable to secondary sinuous modes in the outer shear layers. Consequently,  a subcritical transition to turbulence, i.e., bypass transition, can be already initiated in the favourable-pressure-gradient region ahead of the wave crest.

\end{abstract}


\section{Introduction}\label{sec:intro}
\begin{flushright}
\begin{figure}
\includegraphics{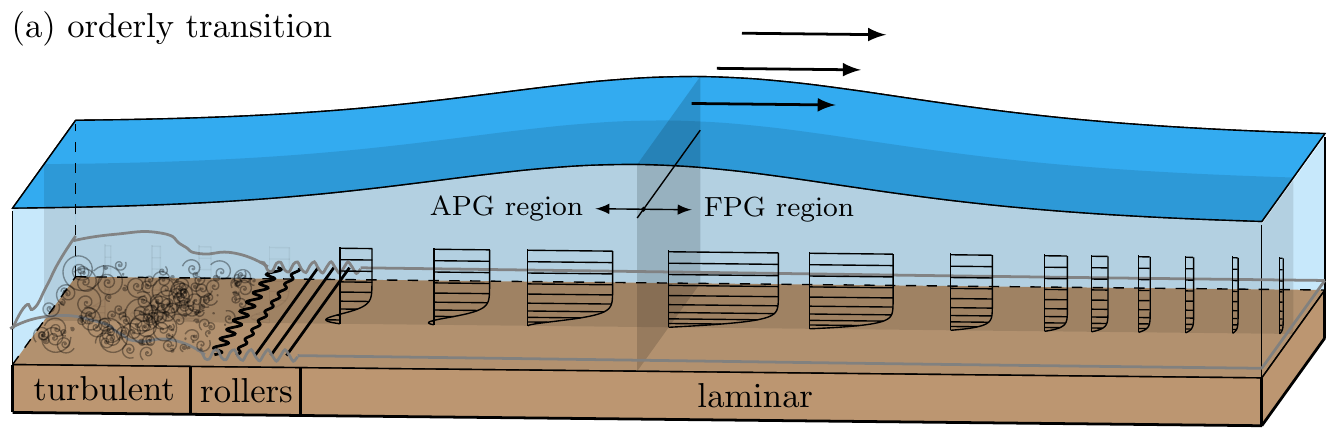}
\vspace{12pt}
\includegraphics{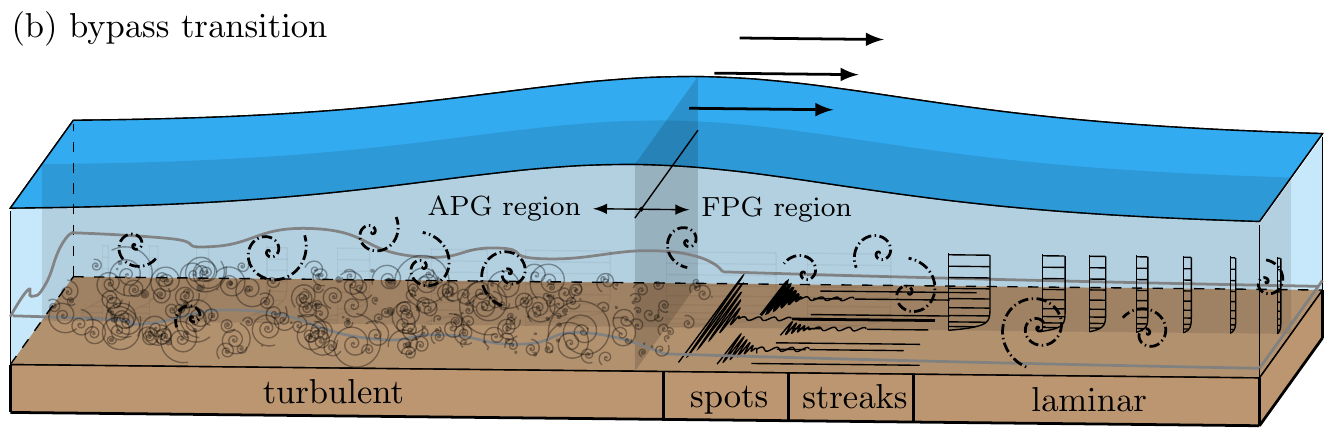}
\caption{\label{fig:sketch} Conceptual sketches showing two main paths of transition to turbulence in the bottom boundary layer under a solitary wave. Scales in the boundary layer are exaggerated for clarity. Laminar velocity profiles are plotted until the onset of transition.  (a) Orderly route to transition via two-dimensional modal instabilities initiated by the inflectional velocity profile. 
(b) Bypass transition initiated by the receptivity of boundary layer to finite amplitude ambient disturbances (dashed-dotted curls). The instability is three-dimensional and of stochastic nature.} 
\end{figure}
\end{flushright}

Solitary waves are long waves of permanent form, which induce approximately constant velocity in the water column \citep{munk1949solitary}. They are subject to friction in the thin boundary layers developing at the free surface and at the sea bottom. The free-surface boundary layer is usually weak \citep{klettner2012laminar}, and is negligible. On the other hand, the bottom boundary layer is of prominent importance, as it hosts the hydrodynamic processes driving the sediment motion and energy dissipation.  In the most basic setting of wave propagating in a constant depth over a smooth bottom, the bottom boundary layer consists of regions of favourable and adverse pressure gradient (FPG and APG) located ahead and behind of the wave crest respectively, cf. figure~\ref{fig:sketch}. The boundary layer flow has tendency to remain laminar in the FPG region. Behind the wave crest, the APG gives rise to an inflectional velocity profile \citep{LIU:2007dv}, cf. velocity profiles in figure~\ref{fig:sketch}a, and the boundary layer becomes linearly unstable  \citep{Blondeaux:2012ei,Verschaeve:2014gh,Sadek:2015jm}. Experimental \citep{Sumer:2010ce} and numerical \citep{Vittori:2008gv,OZDEMIR:2013bu} models of solitary-wave boundary layer (SWBL) have shown that the inflectional instability leads to regularly spaced spanwise-oriented vortex rollers, which can break down to small-scale turbulence in higher wave amplitudes. 

Linearly stable base flow in the FPG region does not preclude the onset of transition in this region. Finite amplitude external perturbations such as breaking-wave turbulence, sound or small-scale bedforms can lead to significant growth in finite times and yield secondary base states that can be unstable. Such subcritical transition can take place in the FPG region before the arrival of the wave, and is analogous to bypass transition in zero-pressure gradient (ZPG) boundary layers \citep{morkovin1969many}, as the modal instability is bypassed by another noise-induced mechanism. This alternative transition scenario is depicted for SWBL in figure~\ref{fig:sketch}b. Unlike the orderly transition, whose initiation is often described simply by a critical Reynolds number, bypass transition is a complicated problem depending on the amplitude, frequency and type of external perturbations in addition to Reynolds number. A recent review on the phenomenology of bypass transition can be found in \cite{DURBIN:2017dn}. All bypass scenarios require a receptivity stage, where external perturbations are modified and amplified by the boundary layer, and a breakdown stage, where the most amplified modes become unstable and break into turbulence. The receptivity stage is dominated by streamwise-oriented vortices and streaks. The former are the surviving modes from rapid shear distortion \citep{phillips1969shear} and the latter are elongated streamwise-momentum modes produced by the former via stirring the base flow, a process known as lift-up mechanism \citep{landahlJFM80,brandt2014lift}. The initial stages of streak amplification can be linked mathematically to the nonnormality of the linarized Navier-Stokes operator  \citep{butlerPOF92,trefethen1993hydrodynamic}. 

The breakdown stage is characterized by secondary instabilities and resultant turbulent spots. Two scenarios have been observed depending on the amplitude of streaks. When the environment forcing is strong, the lift-up mechanism can generate highly elevated low-speed streaks acting like strong wake perturbations on their environment. These protruding layers are susceptible to wake-like instabilities driven by spanwise shear \citep{waleffe1995hydrodynamic,Andersson:2001dm,Vaughan:2011ho}. Consequently, they develop rapidly growing sinuous undulations, and break down into turbulent spots \citep{matsubaraJFM2001,jacobsJFM01,hernonJFM07}.  In contrast, when the environment forcing is modest, the streaks are weaker and remain confined to the near-wall region. In this case, instabilities can occur on vertical shear layers that are slightly modulated by streaks. These instabilities are observed to have reduced growth rates compared to reference instabilities (Tollmien--Schlichting (TS) waves), cf. \cite{cossu2004tollmien} and \cite{liu2008floquet}. Therefore, introducing moderate-amplitude streaks to the boundary layer can delay transition point to turbulence \citep{cossu2002stabilization}. \cite{Vaughan:2011ho} named the two streak instabilities after the location of their respective critical layers and called them ``outer'' and ``inner'' modes. We will adapt a similar terminology for the streak instabilities in the present work.  The final step of breakdown stage follows the same path for both inner and outer modes, i.e., turbulent spots at different locations grow and amalgamate (e.g. \cite{Narasimha:1985dd}), and finally, turbulent boundary layer sets in.

Unlike flat-plate boundary layers, experimental and numerical evidence for bypass transition in SWBLs are sparse.  Using direct numerical simulations (DNS), \cite{OZDEMIR:2013bu} examined the effect of the perturbation amplitude by seeding random noise of varying magnitudes (between $1-20 \%$ of maximum free-stream velocity) to initial conditions before the arrival of the solitary wave. The cases with $5\%$ or more noise and $Re_\delta\geq1500$, where Reynolds number is defined using Stokes length and maximum free-stream velocity (cf. \S~\ref{sec:problem} for details), showed an initial energy amplification inside the boundary layer lasting until another more rapidly growing amplification mechanism takes over in the APG stage after flow reversal.  They speculated that this early perturbation growth should be due to a nonlinear viscous instability, as it takes place in the FPG stage where velocity profiles do not contain any inflection point, a necessary condition for inviscid instability. However, it is more likely that the amplification is due to linear transient nonnormal growth. Indeed,  \cite{Verschaeve:2017fa} found a strong linear nonnormal growth in the FPG stage of SWBL if the initial perturbations are organized as streamwise rollers. These rollers then amplify streaks by the lift-up mechanism with a maximum growth proportional to the square of Reynolds number. Later in the APG stage, the nonnormal growth of streaks are overtaken by the nonnormal growth of two-dimensional spanwise modes, as the maximum growth of these modes has an exponential scaling in Reynolds number. The analysis of \cite{Verschaeve:2017fa} provided conceptual insights for the overtaking of another growth mechanism in the APG stage in \cite{OZDEMIR:2013bu}. However, there is some discrepancy between DNS and transient-growth analysis at which critical Reynolds number the nonnormal two-dimensional modes begins to exert dominance. This is possibly related to nonlinear effects, which are not considered in the study of \cite{Verschaeve:2017fa}. The transition process in \cite{OZDEMIR:2013bu} was initiated by regularly spaced vortex rollers in all cases. Although bypass transition via streak breakdown was not observed, the secondary mechanisms in transition were sensitive to the level of initially seeded perturbation suggesting a sensitivity to the presence of the streaks. Low-noise cases followed a transition path reminiscent of free-shear layers. In contrast, in high-noise cases, where the streaks should be strongest, vortex rollers broke into $\Lambda$-shaped vortices, hence the transition was reminiscent of K-type transition in flat-plate boundary layers \citep{klebanoff1962three}. 

 \cite{Sumer:2010ce} simulated a SWBL in an oscillatory water tunnel and observed turbulent spots in a flow regime starting at $\Rey_\delta=1000$. These sporadic features spread to earlier phases with increasing Reynolds number.  At the highest Reynolds number achieved ($\Rey_\delta=2000$), they were already observed at the end of FPG stage. Turbulent spots grew and merged but did not lead to a complete breakdown in any of the cases. They coexisted concurrently with vortex rollers, which emerged later  in laminar areas surrounding turbulent spots, cf. video~3 in supplemental materials of \cite{Sumer:2010ce}. The precursor structures to these turbulent spots are not clear. It is difficult to discern streamwise streaks in the visualizations and supplemental movies of \cite{Sumer:2010ce}. However, in a prequel paper \citep{Carstensen:2010dl}, where an oscillatory flow was investigated in the same apparatus, it was clearly shown that streamwise streaks are precursors to turbulent spots in transitional oscillatory boundary layers. It is possible that the same facility noise induced streaks in both periodic and solitary wave boundary layers. To date, the streak breakdown in SWBLs is explicitly shown only in the DNS study of \cite{sadek2015mechanisms}, where a bypass scenario is initiated by seeding optimal streamwise-constant perturbations and some localized secondary perturbations towards the end of the FPG stage. The injected streamwise streaks became unstable and the breakdown to small-scale turbulence took place in the APG stage.

Studies by \cite{OZDEMIR:2013bu} and \cite{Verschaeve:2017fa} imply that the FPG stage of SWBL is receptive to environment perturbations and can respond by developing streaks. There is some experimental evidence that these streaks might breakdown into turbulent spots  \citep{Sumer:2010ce} or modify the secondary modes of transition when they have modest energy \citep{OZDEMIR:2013bu}.  Furthermore, we anticipate that modest-amplitude streaks can have a stabilizing effect on the instabilities developing in the APG stage as in flat-plate boundary layers \citep{cossu2004tollmien}. There is a need for a systematic study to determine the quantitative and qualitative extent of these effects. In particular, the receptivity and breakdown stages of bypass transition in SWBLs have to be characterised in more detail. The present study is an effort in this direction. 

The disturbances in the sea environment continuously force the wave boundary layer in a random fashion. To this end, the flow structures dominating the early landscape in the boundary layer are induced by the external perturbations to which the boundary layer shows the strongest response. These perturbations can be identified in an optimisation framework using a system perspective, where external disturbances provide the input and the boundary-layer response corresponds to the output. A convenient approach to model the external perturbations is using body forces of stochastic or deterministic nature. In this context, a deterministic perturbation model allows a more controllable approach, where the frequency and spatial distribution of body forces can be specified. Assuming perturbations have small amplitude, a linear approach can be utilized and response to all possible disturbances in the form of Fourier modes can be investigated.  In a pioneering work using this model, \cite{Jovanovic:2005ij} studied the linear response of the plane Pouseille flow, and identified counter-rotating streamwise-constant rollers as the most ``dangerous'' perturbation delivering the largest amplification per energy input. The rollers induced energetic streamwise-constant streaks via the lift-up mechanism. This study showed that the linear input-output framework can capture the essence of receptivity stage despite its simplicity.  We will use a similar approach as a starting point in the present analysis and study the receptivity of SWBL. We obtain input-output configurations in the form of streak-roller systems similar to \cite{Jovanovic:2005ij}. Subsequently, the breakdown stage is investigated with a combination of linear secondary stability analysis and fully nonlinear numerical simulations triggered with finite amplitude perturbations. We quantify the critical perturbation levels leading to breakdown of streaks via inner and outer secondary instabilities. Interesting results are found implying a dual role for the streaks. Weak to moderate amplitude streaks and associated inner instabilities are shown to have a stabilizing effect on the boundary layer, whereas higher amplitude streaks can lead to an early bifurcation to an unstable branch already in the FPG stage.

The manuscript is organized as follows. In \S\ref{sec:problem}, we briefly introduce the SWBL in a temporal setting. Subsequently, the linear input-output framework is described in \S\ref{sec:Gmax}, where the linear flow response is also discussed.  We then select a suitable excitation configuration and embed it to nonlinear governing equations  in \S\ref{sec:streaks} to obtain streaky SWBLs, which are the flow response to finite amplitude excitation. The nonlinear flow response represents the secondary flow states that are amenable to linear instabilities. The character and phase of these instabilities are analysed in \S\ref{sec:transition} using a linear secondary stability analysis based on a quasi-static assumption. In \S\ref{sec:breakdown}, the growth and breakdown of streaks is studied using nonlinear DNS. The objective of this section is the validation of quasi-static assumption and the determination of breakdown thresholds. Finally,  in \S\ref{sec:conclusion}, the results are summed up, and implications of the analysis are discussed with some outlook for future work.

\section{Problem formulation}\label{sec:problem}

We consider a small-amplitude solitary wave with a wave height $H^*$ propagating over a constant depth $h^*$. In this work, the physical quantities with asterisk are dimensional quantities. The problem is defined in a Cartesian coordinate system, where $x^*$ is the direction of wave propagation (also called streamwise direction), $y^*$ is the spanwise direction parallel to wave crest, and $z^*$ is the vertical direction extending from the bed upwards. The velocity components associated with these directions are $u^*$, $v^*$, and $w^*$, respectively. The leading order solution of  solitary wave profile is given by \citep{grimshaw1970solitary}
\begin{equation}
\frac{\eta_w^*}{h^*}=1+\frac{H^*}{h^*}\mathrm{sech}^2\left(\sqrt{\frac{3H^*}{4h^{*3}}}(x^*-c_w^*t^*)\right),
\end{equation} 
where $c_w^*=\sqrt{g^*h^*}$ is the wave speed with $g^*$ being the gravitational acceleration. Furthermore, the irrotational streamwise velocity, which is constant in the water column, is given by
\begin{equation}
u_0^*(x^*,t^*)=U_{0m}^*\mathrm{sech}^2\left(\sqrt{\frac{3H^*}{4h^{*3}}}(x^*-c_w^*t^*)\right),
\label{eq:velAmbx}
\end{equation}
with the maximum velocity
\begin{equation}
U_{0m}^*=\sqrt{g^*h^*}\frac{H^*}{h^*}.
\end{equation}

Adjacent to the bed,  there is a bottom boundary layer, in which the streamwise velocity $u^*$ has also a rotational velocity component $u_r^*$. The thickness of the boundary layer is usually much smaller than the water depth, cf. Appendix~A in \cite{Sumer:2010ce}. Therefore, streamwise variations in the boundary layer can be considered negligible compared to temporal and vertical variations. These assumptions allow a local parallel formulation of the bottom boundary layer, where the flow is assumed homogeneous in horizontal directions. At a fixed point, the irrotational velocity at the bottom (free-stream velocity hereafter) depends now only on time and reads as follows
\begin{equation}
u_0^*(t^*)=U_{0m}^*\mathrm{sech}^2(-\omega_w^*t^*),
\label{eq:velAmb}
\end{equation}
where
\begin{equation}
\omega_w^*=\sqrt{\frac{3g^*H^*}{4h^{*2}}}
\end{equation}
is the effective wave frequency. Using the wave frequency and kinematic viscosity the Stokes length is defined
\begin{equation}
\delta_s^{*}=\sqrt{{2\nu^{*}}/{\omega_w^{*}}}
\label{eq:Stokes}
\end{equation}
as the boundary-layer scale of the problem. Equation (\ref{eq:velAmb}) neglects the first and higher-order terms in ${H^*}/{h^*}$.  Therefore, the model is relevant only for ${H^*}/{h^*}\rightarrow 0$.  \cite{Vittori:2011da} employed a less restrictive model, in which first- and second-order terms in ${H^*}/{h^*}$ are also included.  The reader is referred to their work for the effect of the wave height on the boundary layer transition under a solitary wave. Assuming ${H^*}/{h^*}\rightarrow 0$ and ${\delta^*}/{h^*}\rightarrow 0$ provides the advantage of reducing the parameter space of the problem to one, the Reynolds number:
\begin{equation}
\Rey_\delta=\frac{U_{0m}^*\delta_s^{*}}{\nu^{*}}.
\label{eq:Re}
\end{equation}
The Stokes length is now the only relevant length scale of the problem. 

\begin{figure}
\begin{center}
\includegraphics{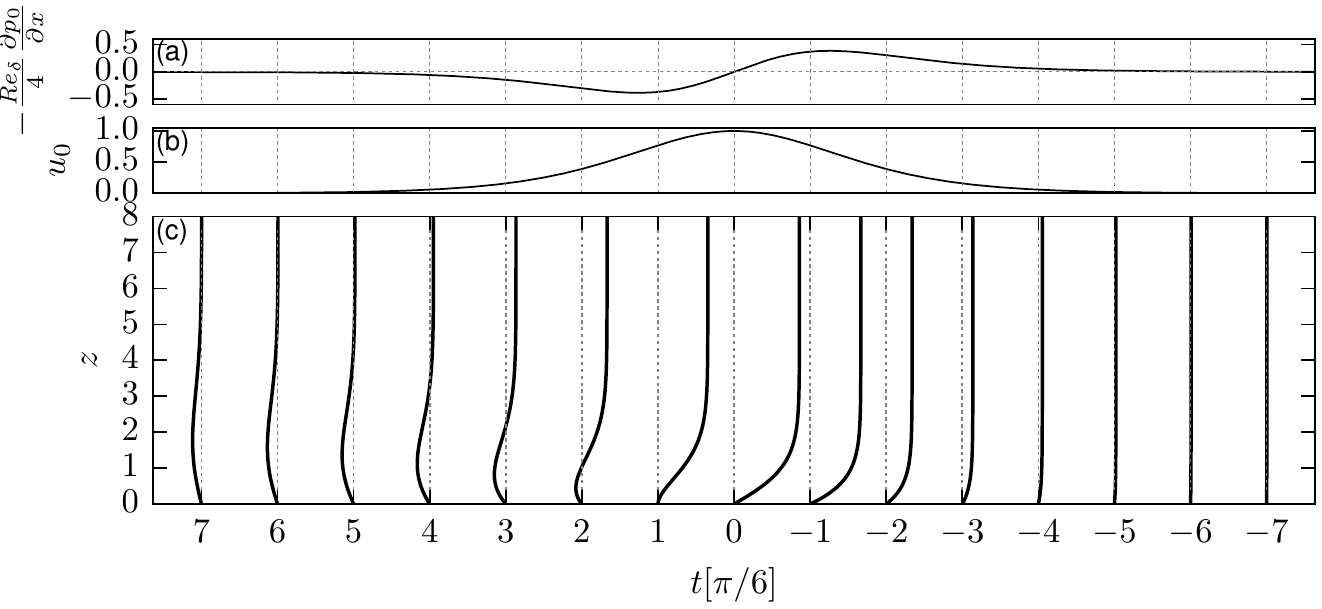}
\caption{(a) Free-stream pressure gradient (\ref{eq:dpdx}); (b) Free-stream velocity (\ref{eq:velAmb}); (c) Vertical profiles of streamwise velocity in a laminar temporal solitary-wave boundary layer. The time axis is scaled with $\pi/6$.}
\label{fig:UBase}
\end{center}
\end{figure}

We introduce the following dimensionless velocity fields, spatial coordinates, time and pressure, respectively,
\begin{equation}
\vec u = \vec u^*/U_{0m}^*;~~\vec x=\vec x^*/\delta_s; ~~t=t^*\omega_w^*; ~~p=p^*/\rho^*U_{0m}^{*2}.
\end{equation}
The momentum equation for the irrotational streamwise velocity at the bottom can be expressed in local temporal frame approximation as
\begin{equation}
\frac{2}{\Rey_\delta}\frac{\mathrm d u_0}{\mathrm d t}=-\frac{\partial p_0}{\partial x},
\label{eq:free}
\end{equation}
The free-stream pressure gradient is calculated using (\ref{eq:velAmb}) and (\ref{eq:free}) and reads
\begin{equation}
-\frac{\partial p_0}{\partial x}=\frac{4}{\Rey_\delta}\mathrm{sech}^2(-t)\mathrm{tanh}(-t).
\label{eq:dpdx}
\end{equation}
The non-dimensional pressure gradient and  free-stream velocity are plotted in figures \ref{fig:UBase}a and \ref{fig:UBase}b, respectively.   The overall balance of streamwise momentum in the laminar SWBL is given by
\begin{equation}
\left(\frac{\partial }{\partial t}-\frac{1}{2}\frac{\partial^2 }{\partial z^2}\right)U=2\mathrm{sech}^2(-t)\mathrm{tanh}(-t),
\label{eq:base}
\end{equation}
where $U=u_r+u_0$ is the total laminar velocity containing both rotational and irrotational components. Equation (\ref{eq:base}) is supplemented with the boundary conditions $U(z=0,t)=0$ and $U(z\rightarrow\infty,t)=u_0$, and we specifiy the initial condition $U(z,-\infty)=0$. The solution of (\ref{eq:base}) is shown in figure \ref{fig:UBase}c. This approximate model of SWBL is  theoretically solved  \citep{LIU:2004jo} and adapted in experimental \citep{Sumer:2010ce,Tanaka:2012cq} and numerical \citep{OZDEMIR:2013bu,Sadek:2015jm,Verschaeve:2017fa} studies.

\section{Optimal disturbances and the flow response}\label{sec:Gmax}
The time-dependent streamwise velocity  $U(z,t)$ in (\ref{eq:base}) is the base state of the problem, which is continuously forced by external perturbations present in the environment. A convenient approach to study the effect of these perturbations is to model them as body forces. In the present parallel flow model, the forcing fields can be defined as the sum of Fourier components
\begin{equation}
[ f_u, f_v, f_w](x,y,z,t)=\iiint_{-\infty}^{\infty} [\hat f_u,\hat f_v,\hat f_w] (z) \mathrm e^{\ii (\alpha x+ \beta y+ \omega_f t)}\mathrm d\alpha \mathrm d\beta d\omega_f,
\end{equation}
where $\alpha$ and $\beta$ are streamwise and spanwise wavenumbers, respectively, and $\omega_f$ is the frequency.  If we consider small-amplitude perturbations, then the flow response to each Fourier component can be studied independently. In this linear regime, the most dangerous flow scenarios can be initiated by finding the Fourier modes inducing the strongest flow response. The objective of this section is to find these Fourier modes using an optimization framework and to analyse the corresponding flow response. In \S~\ref{sec:IO}, we introduce the optimization problem and the adjoint method to find its solution. Subsequently, the optimal input-output configurations and their scalings are discussed in \S~\ref{sec:linearStreaks}. 

\subsection{Methodology}\label{sec:IO}
We apply a Fourier ansatz in the homogeneous $x$ and $y$ directions for the perturbation velocity and pressure
\begin{equation}
\label{eq:uFourier}
[\tilde u,\tilde v,\tilde w,\tilde p](x,y,z,t)=\mathrm {Re}\{ [\hat u,\hat v,\hat w,\hat p] (z,t) \mathrm e^{\ii (\alpha x+ \beta y)}\}.
\end{equation}
In the linear regime, each Fourier mode is excited by the corresponding harmonic force at the same spatial wavenumber 
\begin{equation}
\label{eq:fFourier}
 [ f_u, f_v, f_w](x,y,z,t)=\mathrm {Re}\{[\hat f_u,\hat f_v,\hat f_w] (z) \mathrm e^{\ii (\alpha x+ \beta y+ \omega_f t)}   \}.
\end{equation}
In the present parallel flow model, the perturbation dynamics can be conveniently studied using the forced versions of the Orr-Sommerfeld and Squire (OSS) equations \citep{Schmid:2014cd}. To this end, two different excitation regimes can be defined. First, when $t\ll-\pi$, the base flow is vanishingly small, and the forcing brings the stagnant flow to a periodic state, which is given by the set
\begin{align}
\label{eq:OSSsteady1}
\frac{1}{\Rey_\delta}\left [\left (2\ii\omega_f-\hat\Delta \right)\hat\Delta\right]\hat w_o(z)&= \hat g_w(z),  \\ 
\frac{1}{\Rey_\delta}\left (2\ii\omega_f-\hat\Delta \right) \hat \eta_o(z)&=\hat g_\eta(z) , \\
\hat w_o(0)=\frac{\partial \hat w_o}{\partial z}(0)=\hat \eta_o(0)&=0, \\
\label{eq:OSSsteady4}
\hat w_o(z\rightarrow\infty)=\frac{\partial \hat w_o}{\partial z}(z\rightarrow\infty)=\hat \eta_o(z\rightarrow\infty)&=0,
\end{align}
where $\hat w$ and $\hat \eta$ are the vertical velocity and vertical vorticity  modes associated with  the wavenumber pair ($\alpha,\beta$), $\hat\Delta=k^2-\partial^2/\partial z^2$ represents the semi-discretised Laplacian operator, $k^2=\alpha^2+\beta^2$, and $\hat g_w$ and $\hat g_\eta$ are the external driving forces containing the control variables $\vec {\hat f}=(\hat f_u,\hat f_v,\hat f_w)$
\begin{align}
\hat g_w&=-\ii \alpha \frac{\partial \hat f_u}{\partial z}- \ii \beta \frac{\partial \hat f_v}{\partial z}-k^2 \hat f_w, \\
\hat g_\eta&=\ii \beta  \hat f_u- \ii \alpha  \hat f_v.
\label{eq:control}
\end{align}

When the wave arrives, the flow is no longer periodic, and the perturbation equations during the wave event becomes
\begin{align}
\label{eq:OSS1}
\left [\left (\frac{2}{\Rey_\delta}\frac{\partial  }{\partial t}+\ii\alpha U-\frac{1}{\Rey_\delta}\hat \Delta \right)\hat\Delta-\ii\alpha \frac{\partial^2 {U}}{\partial z^2}\right ]\hat w(z,t)&= \hat g_w(z)\ee^{\ii\omega_f t}, \\ 
\label{eq:OSS2}
\left (\frac{2}{\Rey_\delta}\frac{\partial  }{\partial t}+\ii\alpha U-\frac{1}{\Rey_\delta}\hat\Delta \right) \hat \eta(z,t)+ \ii \beta \frac{\partial U}{\partial z}\hat w(z,t)&=\hat g_\eta(z)\ee^{\ii\omega_f t}, \\
\hat w(z,-\infty)=\hat w_o(z);~~~~ \hat \eta(z,-\infty)&=\hat \eta_o(z),\\
\hat w(0,t)=\frac{\partial \hat w}{\partial z}(0,t)=\hat \eta(0,t)&=0, \\
\label{eq:OSS5}
\hat w(z\rightarrow\infty,t)=\frac{\partial \hat w}{\partial z}(z\rightarrow\infty,t)=\hat\eta(z\rightarrow\infty,t)&=0. 
\end{align}
The following compact notation is used for periodic and temporal OSS system of equations
\begin{equation}
L_o \vec {\hat q_o}=C\vec {\hat f},~~~~L(t) \vec {\hat q}=C\vec {\hat f}\ee^{\ii\omega_f t}
\end{equation}
where $\hat {\vec q}_o=[\hat w_o, \hat \eta_o]$ and $\hat {\vec q}=[\hat w, \hat \eta]$.  

The response of the flow to an excitation at a wavenumber pair $(\alpha,\beta)$ is measured by the perturbation kinetic energy
\begin{equation}
\label{eq:Euvw}
E(\vec {\hat u} ):=\frac{1}{2}\intz (|\hat u|^2+|\hat v|^2+|\hat w|^2)\mathrm dz.	
\end{equation}
 We can express $\hat u$ and $\hat v$ in terms of $\hat w$ and $\hat\eta$, and obtain  \citep{Schmid:2001hj}
\begin{align}
E(\hat{\vec q}):=\frac{1}{2k^2}\intz (k^2|\hat w|^2+|\frac{\partial \hat w}{\partial z}|^2+|\hat \eta|^2)\mathrm dz.
\label{eq:E}
\end{align}

We look for the most dangerous perturbations initiating the strongest response in the linear SWBL. This is equivalent to finding an optimal control $\vec {\hat f}^{opt}(z;\alpha,\beta,\omega_f,T_f,\Rey_\delta)$, which yields the maximum energy amplification at a terminal time $t=T_f$ per initial energy input $E(\vec{\hat  q_o})$. This is found by solving the constrained optimization problem
\begin{align}
\label{eq:GOpt}
\begin{split}
G_f(\alpha,\beta,\omega_f,T_f,\Rey_\delta)&:=\max\limits_{ \vec {\hat f}}\frac{E(\vec {\hat q}(T_f))}{E(\vec{\hat  q_o})} \\ 
& s.t.\\ 
L_o \vec {\hat q}_o&=C\vec {\hat f}, \\
L(t) \vec {\hat q}&=C\vec {\hat f}\ee^{\ii\omega_f t}, \\
\| \vec {\hat f}\|&=1
\end{split}
\end{align}
where $G_f$  is the largest response or gain. The optimization problem (\ref{eq:GOpt}) is subject to constraints in the form of periodic and transient OSS systems, and to an additional normalization constraint, which ensures the forcing energy is unity.  This optimal control analysis is closely related to the optimal transient-growth analysis of \cite{Verschaeve:2017fa} but differs in control variables, i.e., instead of the growth of initial perturbations, the response to external forcing is measured.

The optimization problem (\ref{eq:GOpt}) is solved using an adjoint approach \citep{Luchini:2014fv}. In this method, a Lagrangian functional is assigned to the optimization problem, and the optimality conditions are derived from the stationary point of the Lagrangian, cf. Appendix~\ref{app:f} for details. To this end, the gain $G_f$ is maximum when the flow is forced by the optimal forcing configuration $\vec {\hat f} ^{opt}$ satisfying
\begin{align}
\nonumber
L_o \vec {\hat q}_o&=C\vec {\hat f}^{opt}, \\
\nonumber
L(t) \vec {\hat q}&=C\vec {\hat f}^{opt}\ee^{\ii\omega_f t}, \\
\label{eq:adjShort}
L^+(t) \vec {\hat q}^+&=0, \\
\nonumber
\| \vec {\hat f}^{opt}\|&=1,\\
 \vec {\hat f}^{opt}&=\rho(\vec{\hat q}^+)
 \label{eq:design}
\end{align} 
Equation (\ref{eq:adjShort}) represents the following adjoint Orr--Sommerfeld and Squire equations
\begin{align}
\label{eq:adjOSS1}
\left [\left (\frac{2}{\Rey_\delta}\frac{\partial  }{\partial t}+\ii\alpha U-\frac{1}{\Rey_\delta}\Delta \right)\Delta-2\ii\alpha \frac{\partial {U}}{\partial z}\frac{\partial}{\partial z}\right ]\hat w^+(z,t)&=-\ii \beta \frac{\partial U}{\partial z}\hat \eta^+(z,t), \\ 
\left (\frac{2}{\Rey_\delta}\frac{\partial  }{\partial t}+\ii\alpha U-\frac{1}{\Rey_\delta}\Delta \right) \hat \eta^+(z,t)&= 0 , \label{eq:adjOSS2}\\
\hat w^+(z,T_f)=-\frac{1}{2k^2}\frac{\hat w(z,T_f)}{E(\vec {\hat q}_o)},~~~\hat \eta^+(z,T_f)&=\frac{1}{2k^2}\frac{\hat \eta(z,T_f)}{E(\vec {\hat q}_o)},\\
\hat w^+(0,t)=\frac{\partial \hat w^+}{\partial z}(0,t)=\hat \eta^+(0,t)&=0, \\
\label{eq:adjOSS5}
\hat w^+(z\rightarrow\infty,t)=\frac{\partial \hat w^+}{\partial z}(z\rightarrow\infty,t)=\hat\eta^+(z\rightarrow\infty,t)&=0. 
\end{align}
The reader is directed to \cite{Schmid:2001hj} for a thorough derivation of equations (\ref{eq:adjOSS1}) and (\ref{eq:adjOSS2}). Furthermore, (\ref{eq:design}) corresponds to the expressions to calculate the optimal forcing configuration using the adjoint fields, i.e.,
\begin{align}
\label{eq:fu}
\hat f_u^{opt}(z)&=-\frac{1}{2\sigma}\intt \left(\ii \alpha \frac{\partial \hat w^+}{\partial z} + \ii \beta\hat\eta^+\right) \ee^{-\ii\omega_f t}  \mathrm dt,\\ 
\hat f_v^{opt}(z)&=\frac{1}{2\sigma}\intt \left(-\ii \beta \frac{\partial \hat w^+}{\partial z} + \ii \alpha\hat\eta^+\right) \ee^{-\ii\omega_f t} \mathrm dt,\\
\hat f_w^{opt}(z)&=-\frac{1}{2\sigma}\intt k^2 \hat w^+  \ee^{-\ii\omega_f t}\mathrm dt,
\label{eq:fw}
\end{align} 
where $\sigma$ is a Lagrange multiplier, cf. Appendix \ref{app:f} for the details of the derivation of (\ref{eq:fu})--(\ref{eq:fw}).

The optimization problem in (\ref{eq:GOpt}) is now transformed to a set of equations with  (\ref{eq:OSSsteady1})--(\ref{eq:OSSsteady4}) and (\ref{eq:OSS1})--(\ref{eq:OSS5}) being the state or forward equations,  (\ref{eq:adjOSS1})--(\ref{eq:adjOSS5}) being the adjoint equations, and (\ref{eq:fu})--(\ref{eq:fw}) being the design equations. These equations are solved in a sequential fashion using a simple adjoint-looping algorithm \citep{Andersson:1999ej}. The algorithm starts with an initial guess of $\vec {\hat f}^{opt}$ and iterates over the following successive steps: (i) calculation of $\vec{\hat q}_o$ using (\ref{eq:OSSsteady1})--(\ref{eq:OSSsteady4}); (ii) forward-in-time integration of the state equations (\ref{eq:OSS1})--(\ref{eq:OSS5}); (ii) backward-in-time integration of the adjoint equations in (\ref{eq:adjOSS1}) and (\ref{eq:adjOSS5}); (iii) updating the control terms with the available adjoint fields using (\ref{eq:fu})--(\ref{eq:fw}) and $\| \vec {\hat f}^{opt}\|=1$. 

The forward and adjoint equations are discretized in space using a spectral method based on Chebyshev polynomials. In this method, the equations are mapped to the domain $\xi\in[-1,1]$ and the Gauss--Lobatto collocation technique is utilized to obtain the discrete set of equations. This is implemented using the differentiation matrices developed by \cite{Weideman:2000hq}. Converged results are obtained for a domain size $z\in[0,20]$ and resolution of $N_z=61$ Chebyshev collocation points in the vertical direction.  The initial time to start to simulations is selected to be $t_i=-10\pi$. At this phase the effect of the wave is negligible. The Crank--Nicolson scheme is employed for time integration.  The time step size is $\delta t=\pi/480$.  A sensitivity analysis confirmed that the selected spatial and temporal resolutions are sufficient.   

\subsection{Linear response of the flow}\label{sec:linearStreaks}
\begin{figure}
\begin{center}

\includegraphics{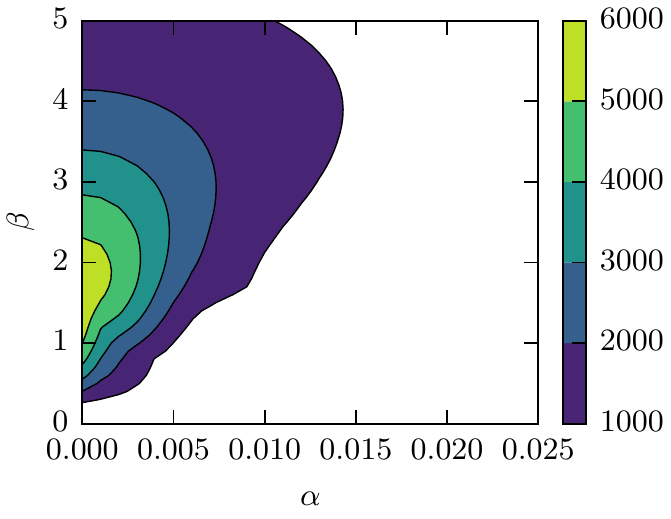}

\caption{Contours of maximum response $G_f(\alpha,\beta,\omega_f=\omega_f^m,T_f=0,\Rey_\delta=2000)$, cf. (\ref{eq:GOpt}), with respect to spanwise wavenumbers and terminal phase for optimization, where $\omega_f^m$ is the excitation frequency delivering the maximum gain. }
\label{fig:Gmax}
\end{center}
\end{figure}

In this section, we study the linear response of the flow to the optimal perturbations at different $\alpha$, $\beta$, $\omega_f$, $T_f$ and $\Rey_\delta$.  Figure~\ref{fig:Gmax} shows the maximum gain among all excitation frequencies for each wavenumber pair $(\alpha,\beta)$ at a moderate Reynolds number $\Rey_\delta=2000$. $\Rey_\delta=2000$ is the highest Reynolds number achieved in the oscillating water tunnel of \cite{Sumer:2010ce}, where they observed turbulent spots at the end of FPG stage. In order to study the receptivity of SWBL among the FPG stage, the terminal time is selected to be $T_f=0$.  It is observed in figure~\ref{fig:Gmax}  that SWBL is very receptive to streamwise-constant ($\alpha=0,\beta\neq0$) excitation in the FPG stage and has a very weak response to two-dimensional $\alpha\neq0,\beta=0$ and oblique $\alpha\neq0,\beta\neq0$ excitations. These modes, mainly two-dimensional ones, become only dominant in mid to late APG stage with the flow reversal. Therefore, they do not play a role in an early subcritical bypass transition, and will not be discussed in the remainder of the text. 

\begin{figure}
\begin{center}
\includegraphics{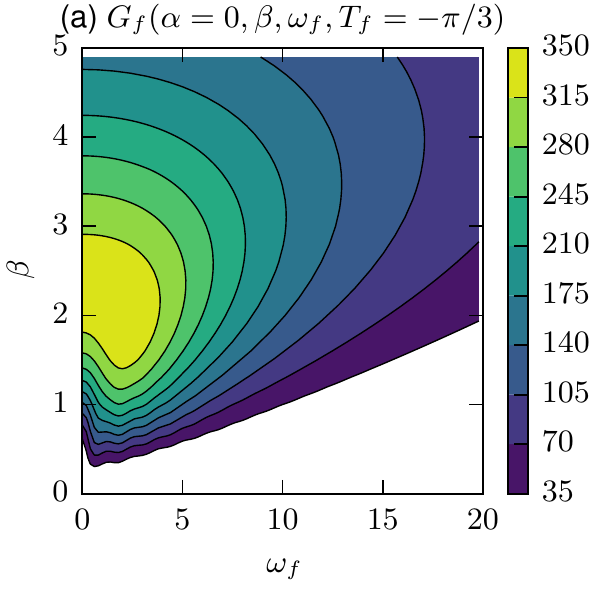}
\includegraphics{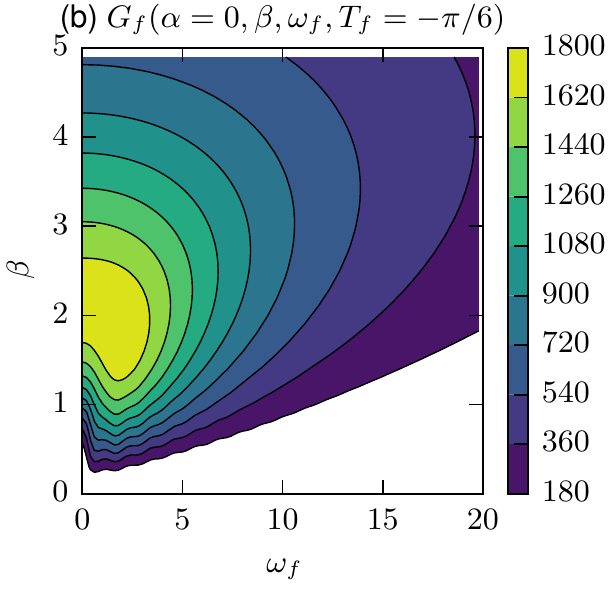}
\includegraphics{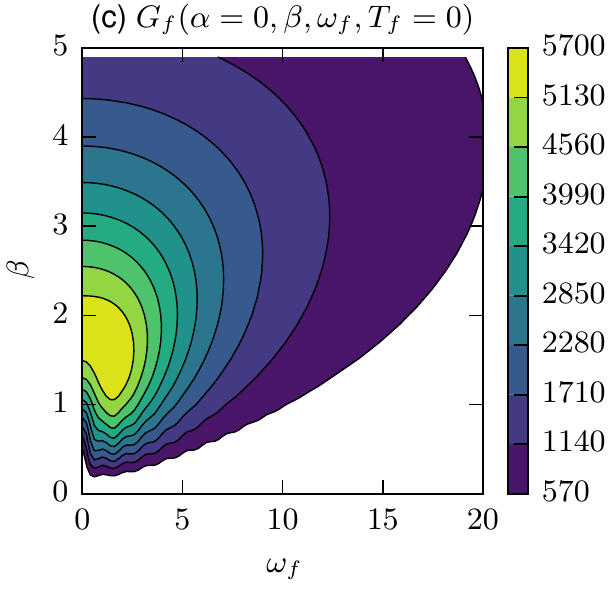}
\includegraphics{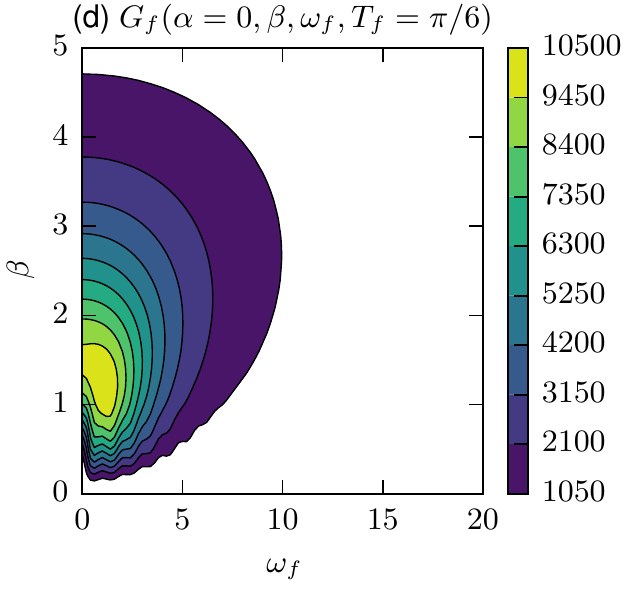}
\caption{Contours of  gain $G_f(\alpha=0,\beta,\omega_f,T_f,\Rey_\delta=2000)$ with respect to spanwise wavenumbers and excitation frequency.  (a) $T_f=-\pi/3$; (b) $T_f=-\pi/6$; (c) $T_f=0$; (d) $T_f=\pi/3$. }
\label{fig:Gmaxa0}
\end{center}
\end{figure}

Figure~\ref{fig:Gmaxa0}  further demonstrates the response of the flow to streamwise-constant excitation on a $\beta-\omega_f$ plane at several terminal times ($T_f=-\pi/3,-\pi/6,0$, and $\pi/6$) at $\Rey_\delta=2000$. We see that with increasing terminal time the frequency band to which the flow is sensitive narrows down. Similarly, there is a shift to lower wavenumbers, which can be linked to the growth of the boundary layer (cf. figure~\ref{fig:UBase}). In general, there is a good flow response in $\beta\in [1.5,2.5]$ and $\omega_f \in [0,3]$. In this range, the boundary layer amplifies the external disturbances up to about $10^4$ times from the start of the wave event   until a phase at the start of APG stage ($T_f=\pi/6$), cf. figure~\ref{fig:Gmaxa0}d. 

In order to analyze the scaling of the governing equations with Reynolds number in the case of streamwise-constant excitation, we introduce the transformations
\begin{equation}
\label{eq:scalings}
\overline {\overline \eta}=\frac{\hat \eta}{\Rey_\delta^2},~~~\overline w=\frac{\hat w}{\Rey_\delta},
\end{equation}
and substitude to streamwise-constant OSS equations (\ref{eq:OSS1}) and  (\ref{eq:OSS2}), where the terms with $\alpha$ vanish, i.e., 
\begin{align}
\label{eq:OSS1a0}
2\left (\frac{\partial  }{\partial t}-\frac{1}{2}\hat \Delta \right)\hat\Delta \overline w&= (- \ii \beta \frac{\partial \hat f_v}{\partial z}-\beta^2 \hat f_w)\ee^{\ii\omega_f t}, \\ 
\label{eq:OSS2a0}
2\left (\frac{\partial  }{\partial t}-\frac{1}{2}\hat\Delta \right) \overline{\overline \eta}&=\frac{1}{\Rey_\delta} \ii \beta  \hat f_u\ee^{\ii\omega_f t}- \ii \beta \frac{\partial U}{\partial z}\overline w. 
\end{align}
In the scaled streamwise-constant setting, the velocity components becomes
\begin{equation}
	\overline{\overline u}=\frac{\ii}{\beta} \overline{\overline \eta}, ~~~ \overline v=\frac{\ii}{\beta}\frac{\partial {\overline w}}{\partial z}.
\end{equation}
Therefore, the evolution of cross-stream momentum by the velocity components $v$ and $w$ is embedded into (\ref{eq:OSS1a0}) and the evolution of streamwise momentum by $u$ is linked to (\ref{eq:OSS2a0}).  As shown in (\ref{eq:OSS1a0}) in the streamwise-constant setting the cross-stream momentum completely decouples from the streamwise momentum. Thus, the cross-stream perturbations are not influenced by the base flow and the wave has no effect on them. The lack of interaction with the wave results in the transverse components remain in their initial state, i.e., 
 \begin{equation}
 	\hat v=\hat v_o\ee^{\ii\omega_f t},~~~\hat w=\hat w_o\ee^{\ii\omega_f t}.
 \end{equation}
 Therefore, the increase in $G_f$ is solely due to intrinsic amplification of $\hat u$ by the boundary layer.

\begin{figure}
\begin{center}
\includegraphics{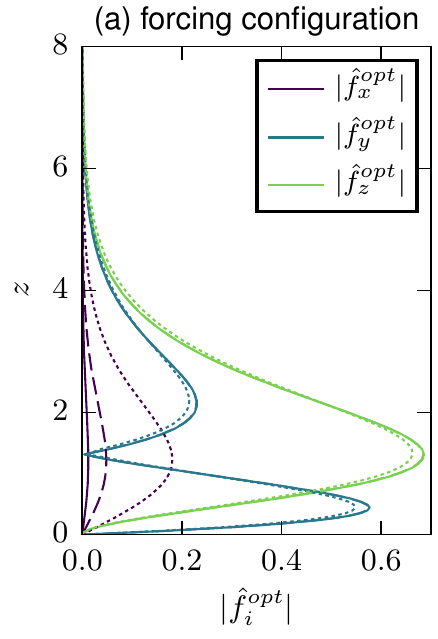}
\includegraphics{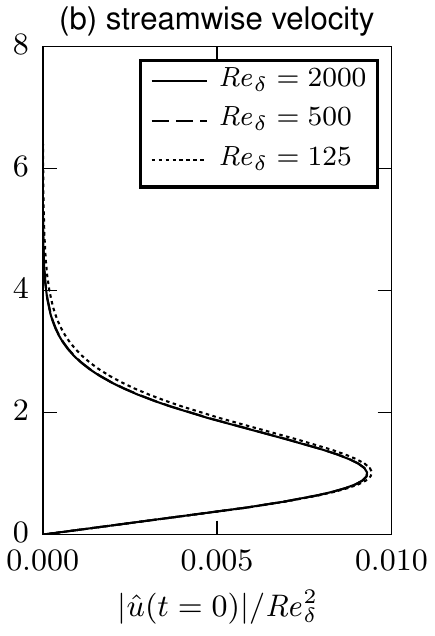}
\includegraphics{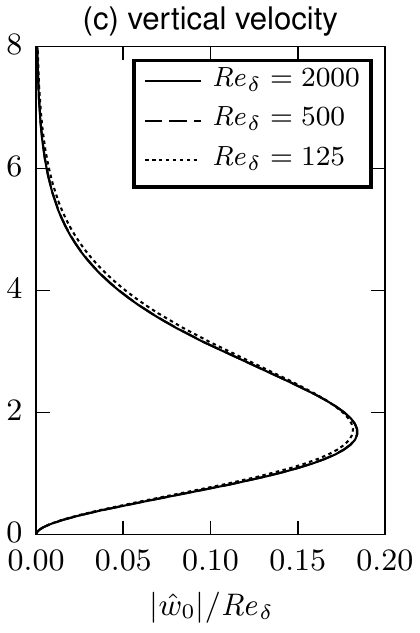}
\end{center}
\caption{\label{fig:fuw} Reynolds-number dependency of the optimal linear input and output fields. (a) Components of the optimal forcing $\vec f^{opt}(\alpha=0,\beta=1.5,\omega_f=0,T_f=0)$ at ($\cdots$): $\Rey_\delta=125$; ($--$): $\Rey_\delta=500$; ($-$): $\Rey_\delta=2000$. See legends for the color coding of forcing components. (b) Streamwise velocity $|\hat u(t=0)|/\Rey_\delta^2$ at the terminal time $t=T_f=0$. (c) Vertical velocity $|\hat w_o|/\Rey_\delta$, which is steady under steady forcing.}
\end{figure}

Equation (\ref{eq:OSS1a0}) suggests that introducing $\overline w$ rendered the cross-stream momentum balance independent of Reynold number,  while (\ref{eq:OSS2a0}) indicates that the streamwise forcing is one order lower ($O(1/Re_\delta)$) than the other $O(1)$ terms . Therefore, in high Reynolds numbers, direct streamwise forcing is inefficient, and optimal external force should concentrate driving cross-stream components, i.e.,  
\begin{equation}
\|\hat f_v^{opt}\|\approx \|\hat f_w^{opt}\|\gg \|\hat f_u^{opt}\|. 
\end{equation}
Consequently, the streamwise forcing in (\ref{eq:OSS2a0}) can be neglected for $\Rey_\delta\gg1$, and the evolution of $\overline{\overline u}$ becomes independent of Reynolds number. Figure~\ref{fig:fuw} validates these Reynolds-number scalings using the numerical results for the case $T_f=0,\alpha=0,\beta=1.5$, and $\omega_f=0$. Similar results are also applicable to other cases. As displayed in figure~\ref{fig:fuw}a  the streamwise component of the optimal force is smaller than the transverse components and it vanishes with increasing Reynolds number. Therefore,  the terminal streamwise velocity scaled with $Re_\delta^2$ and the terminal vertical velocity scaled with $Re_\delta$ collapse for different Reynolds numbers,  cf. figures~\ref{fig:fuw}b and \ref{fig:fuw}c. 

\begin{figure}
\begin{center}
\includegraphics{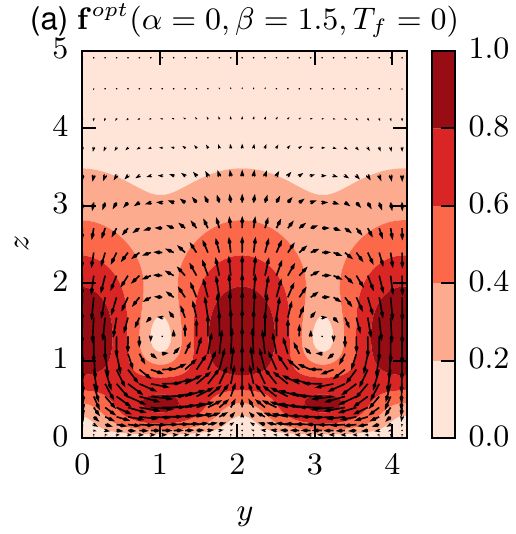}
~~~~~~\includegraphics{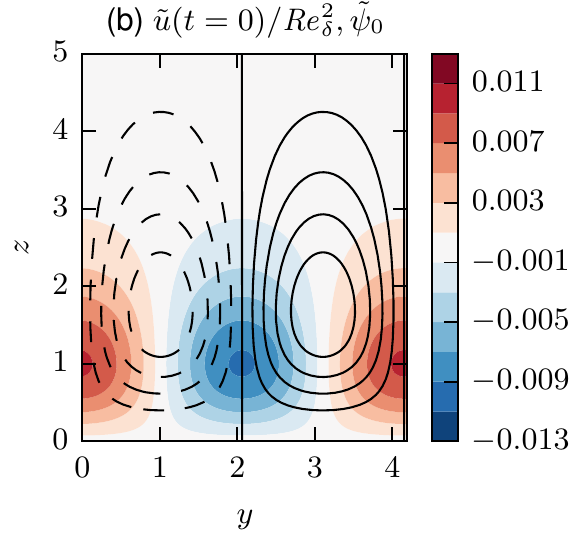}

\end{center}
\caption{\label{fig:streakRoller} Optimal linear input and output configurations in the physical space. (a)  Cross-stream components of the optimal steady streamwise-constant force $\vec f^{opt}(\alpha=0,\beta=1.5,\omega_f=0,T_f=0,\Rey_\delta=2000)$. Filled contours show the forcing magnitude $|\vec f^{opt}|/\max\{\vec f^{opt}\}$. Streamwise component is negligible. Arrows show $f^{opt}_v\hat {\vec j}+f^{opt}_w\hat {\vec k}$, where $\hat{\vec j}$ are $\hat{\vec k}$ are the Cartesian unit vectors in spanwise and vertical directions, respectively. This is the forcing configuration employed for the analysis in \S4-6. (b) The flow response at the terminal time $t=T_f=0$.  Filled contours show levels of streamwise component $\tilde u/\Rey_\delta^2$ and line contours show the steady streamfunction $\tilde \psi_o/\Rey_\delta$ spanning 9 levels between minimum and maximum values in the plane. }
\end{figure}

We now turn to input and output configurations. The optimal steady streamwise-constant forcing configuration $\vec f^{opt}(\alpha=0,\beta=1.5,\omega_f=0,T_f=0)$ and the resulting flow response at the terminal time are shown in the physical space in figures~\ref{fig:streakRoller}a and \ref{fig:streakRoller}b, respectively. In figure~\ref{fig:streakRoller}b, the contour lines present the streamfunction defined as follows
\begin{equation}
\label{eq:psi}
\tilde v=-\frac{\partial \tilde \psi}{\partial z}, ~~~\tilde w=\frac{\partial \tilde \psi}{\partial y}.
\end{equation}
It is observed that the steady forcing is organized as counter-rotating cells $(0,\hat f_v^{opt},\hat f_w^{opt})$, which induce steady rollers  $(0,\tilde v_o,\tilde w_o)$ with the same sense of rotation. The rollers redistribute the streamwise momentum of the base flow, while they lift up the low momentum fluid and pull down the high momentum fluid. As a result, streaks that are antiphase, with the vertical velocity are produced, i.e., regions of negative $\tilde u$ and positive $\tilde w_o$, and vice versa, collapse. There is no feedback from streaks to rollers, as long as the streaks remain stable. We will see later that the same observation also applies to nonlinear streamwise-constant equations. Streaks are merely forced by the linear interaction between the base flow and vertical perturbations. In (\ref{eq:OSS1a0}) and (\ref{eq:OSS2a0}), $\overline {\overline \eta}$ and $\overline {w}$ are of the same order in Reynolds number. So, a transverse steady forcing with an amplitude of $O(1/Re_\delta^2)$ will induce steady rollers of strength $O(1/Re_\delta)$. These rollers then in turn force the streaks of $O(1)$, which will grow with increasing rates associated with the outer-velocity time-scales. \cite{waleffe1995hydrodynamic} derived a similar streak--roller system, where $O(1)$ streaks synthetically forced by the steady rollers of magnitude $O(1/Re_\delta)$. These scalings in Reynolds number also apply to the streaks forced by optimal initial perturbations in the form of counter-rotating rollers in steady boundary layers \citep{Gustavsson:1991fe,Schmid:2001hj} and in SWBLs \citep{Verschaeve:2017fa}.

 \section{Nonlinear streaks} \label{sec:streaks}
When the perturbations reach appreciable amplitudes, nonlinear effects should be taken into account.  We showed in \S~\ref{sec:linearStreaks} that the cases with $\beta\neq0$, $\alpha=0$ and $\omega_f=0$ present a good balance between the optimality and simplicity. Therefore, hereafter the discussion will focus on optimal steady streamwise-constant perturbations, which are arranged as streaks and rollers.  In this configuration, the forcing concentrates in cross-stream components and induces rollers that remain steady also in nonlinear regimes due to lack of interaction with the wave. Therefore, the velocity field of the nonlinear rollers is found by 
\begin{align}
\label{eq:nonlinV}
 -\frac{1}{\Rey_\delta}\Delta \tilde v_o&= -\frac{\partial \tilde p_o}{\partial y}+A_of_v^{opt}-\left(\tilde v_o\frac{\partial \tilde v_o}{\partial y}+\tilde w_o\frac{\partial \tilde v_o}{\partial z}\right), \\ 
\label{eq:nonlinW}
-\frac{1}{\Rey_\delta}\Delta \tilde w_o&= -\frac{\partial \tilde p_o}{\partial z}+A_of_w^{opt}-\left( \tilde v_o\frac{\partial \tilde w_o}{\partial y}+\tilde w_o\frac{\partial \tilde w_o}{\partial z}\right), \\
\label{eq:nonlinCon}
\frac{\partial \tilde v_o}{\partial y} +\frac{\partial \tilde w_o}{\partial z}&=0,
\end{align}
where $A_o$ is a small forcing magnitude with $A_o\ll1$ and $\Delta$ is the Laplacian operator. The steady rollers excite the streaks via intermodal nonlinear interactions and also  via linear interaction with the base flow, i.e., 
\begin{align}
\label{eq:nonlinU}
\frac{2}{\Rey_\delta}\left(\frac{\partial  }{\partial t}-\frac{1}{2}\Delta\right) \tilde u &=
-\left(\tilde v_o\frac{\partial \tilde u}{\partial y}+\tilde w_o\frac{\partial \tilde u}{\partial z}+\tilde w_o\frac{\partial U}{\partial z}\right),
\end{align}
where the small streamwise forcing $A_of_u^{opt}$ is neglected. As the evolution of the cross-stream momentum remains decoupled from the streamwise momentum also in the nonlinear regime, there is no feedback from nonlinear streaks to rollers as long as streaks go through streamwise-constant deformations, which is the case for the stream-constant excitation. More generic perturbations are to be introduced in \S~\ref{sec:transition}. Before proceeding with the results for this nonlinear streak--roller system, we first transform the nonlinear equations to a more convenient form with the aim of reducing the number of parameters in the analysis. To this end, we introduce the variable 
\begin{equation}
\label{eq:A}
A=A_o\Rey_\delta^2,
\end{equation}
and define the transformations
\begin{equation}
\dbtilde v,\dbtilde w,\dbtilde p=\frac{\Rey_\delta}{A}\tilde v,~\frac{\Rey_\delta}{A}\tilde w,~\frac{\Rey_\delta^2}{A}\tilde p.
\end{equation}
Introducing the transformed variables to the roller equations
\begin{align}
\label{eq:nonlinVs}
 -\Delta \dbtilde v_o&= -\frac{\partial \dbtilde p_o}{\partial y}+f_v^{opt}-A\left(\dbtilde v_o\frac{\partial \dbtilde v_o}{\partial y}+\dbtilde w_o\frac{\partial \dbtilde v_o}{\partial z}\right), \\ 
\label{eq:nonlinWs}
-\Delta \dbtilde w_o&= -\frac{\partial \dbtilde p_o}{\partial z}+f_w^{opt}-A\left( \dbtilde v_o\frac{\partial \dbtilde w_o}{\partial y}+\dbtilde w_o\frac{\partial \dbtilde w_o}{\partial z}\right), \\
\label{eq:nonlinCons}
\frac{\partial \dbtilde v_o}{\partial y} +\frac{\partial \dbtilde w_o}{\partial z}&=0,
\end{align}
and to the streak equation
\begin{align}
\label{eq:nonlinUs}
\left(\frac{\partial  }{\partial t}-\frac{1}{2}\Delta\right) \tilde u &=
-\frac{A}{2}\left(\dbtilde v_o\frac{\partial \tilde u}{\partial y}+\dbtilde w_o\frac{\partial \tilde u}{\partial z}+\dbtilde w_o\frac{\partial U}{\partial z}\right),
\end{align}
Transforming the nonlinear governing equations from (\ref{eq:nonlinV})--(\ref{eq:nonlinU}) to  (\ref{eq:nonlinVs})--(\ref{eq:nonlinUs}) reduces the parameter space of the problem from two, $\Rey_\delta$  and $A_0$, to one,  $A$, which can be considered now as the effective amplitude of the excitation. We reiterate that this one-parameter model is only applicable in the range of $\Rey_\delta\gg1$, where optimal forcing configuration $\vec f^{opt}$ does not depend on $\Rey_\delta$ and has a vanishing streamwise component.

The streak--roller equations are solved using the open-source CFD library Nektar++ \citep{cantwell2015nektar++}. To this end, a high-order spectral element method is employed in a two-dimensional computational domain extending to $z\in[0,L_z=20]$ in the vertical direction, and to $y\in[0,Ly=2\pi/\beta]$ in the spanwise direction. Periodicity is applied in the $y$ direction.
The domain discretized using a structured two-dimensional grid with $N_y=24$ and $N_z=36$ elements in spanwise and vertical directions, respectively. The grid is clustered towards to wall, and the expansion rate of elements in the vertical direction is set to $1.1$. Each element is equipped with two dimensional nodal expansion bases, which are constructed using Lagrange polynomials  that are defined on Gauss--Lobatto--Legendre points \citep{karniadakis2005spectral}. A polynomial order of $N_p=7$ is employed. The governing equations are projected on the polynomial basis using a continuous Galerkin method. The resulting system of differential algebraic equations is discretized in time using an implicit second-order scheme, cf. \cite{vos2011generic} for details. Finally, the coupled linear system of equations is segregated using a velocity-correction scheme \citep{karniadakis1991high}. 

\begin{figure}
\begin{center}
\includegraphics{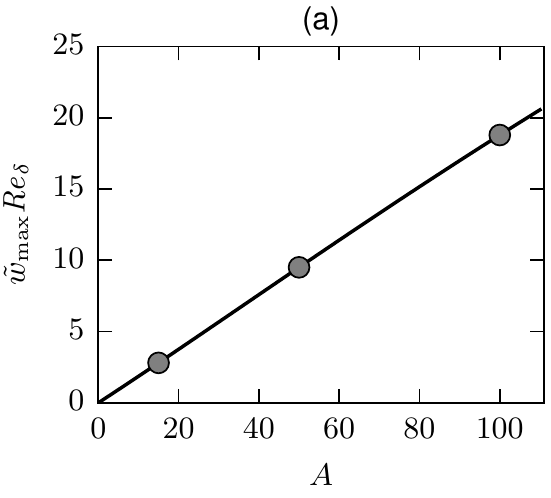}
~~~~~~~~~~\includegraphics{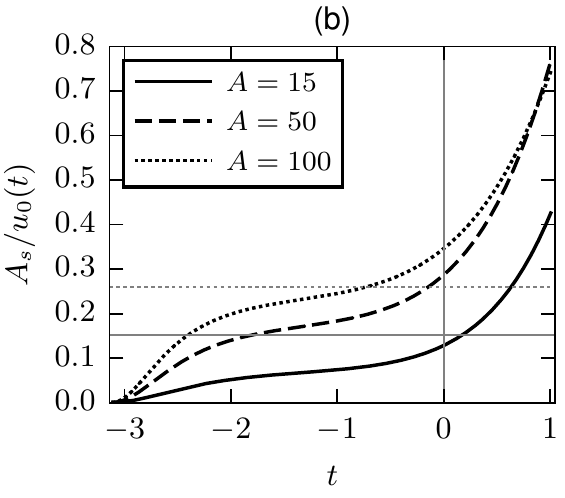}
\end{center}
\caption{\label{fig:wMax} (a) Variation of the roller amplitudes (\ref{eq:wMax}) with respect to the forcing amplitude A (\ref{eq:A}) in the case of linearly optimal forcing with $\alpha=0,\beta=1.5, \omega_f=0$, and $T_f=0$. The symbols mark the values at $A=15,50$~and~$100$, for which the evolution of streak amplitudes (\ref{eq:As}) is shown in (b). The spatial distribution corresponding to the forcing with these amplitudes are presented in figure~\ref{fig:nonStreaks15}. The light horizontal lines in (b) shows two different critical streak amplitudes that are reported in the literature for the emergence of instabilities on steady streaks. (solid line): $A_s^c=0.152$ by \cite{Vaughan:2011ho}; (dashed line): $A_s^c=0.26$ by \cite{Andersson:2001dm}.}
\end{figure}

To keep the analysis on the evolution, stability and breakdown of nonlinear streaks in a tractable margin, a selection has to be made for a representative spanwise wavenumber $\beta$ and terminal time $T_f$. To this end, $T_f=0$ is a good choice to obtain strong amplification during the FPG stage. Furthermore, we see in figure~\ref{fig:Gmaxa0}c,d that the wavenumber $\beta=1.5$ shows good performance for time horizons corresponding to the strongest amplifications ($T_f=0,\pi/6$). Therefore, we will merely consider roller perturbations induced by optimal forcing $\vec f^{opt}(\alpha=0,\beta=1.5,\omega_f=0,T_f=0)$ in the current and upcoming sections. This forcing configuration was shown in figure~\ref{fig:streakRoller}a above.

It is convenient to characterize the nonlinear streaks and rollers via simple scalar measures for their amplitudes. Following \cite{Andersson:2001dm}, amplitude of streaks is defined as the half of the difference between maximum and minimum perturbation velocity, i.e., 
\begin{equation}
\label{eq:As}
A_s(t)=\frac{1}{2}\left (\max_{y,z}\{\tilde u(y,z,t)\}-\min_{y,z}\{\tilde u(y,z,t)\}\right). 	
\end{equation}
$A_s$ approaches to the peak of Fourier mode ($\hat u$) in the linear regime. The amplitude of steady rollers can be prescribed conveniently using the maximum vertical velocity, i.e., 
\begin{equation}
\label{eq:wMax}
\tilde w_{\max}=\max_{y,z}\{\tilde w_o(y,z)\}. 	
\end{equation}
Figure~\ref{fig:wMax}a shows the variation of the roller magnitudes  with respect to effective forcing amplitude $A$. The amplitudes are presented in a $\Rey_\delta$-independent scaling, i.e., $\tilde w_{\max}\Rey_\delta=A\dbtilde w_{\max}$. We see that the rollers are in approximately linear regime for the considered range of forcing amplitudes $A$.  Figure~\ref{fig:wMax}b further shows the temporal evolution of normalized streak amplitudes $A_s/u_0(t)$ for the cases $A=15,50$~and~$100$ corresponding to roller magnitudes $\tilde w_{\max}=2.8/\Rey_\delta,9.51/\Rey_\delta$ and $18.78/\Rey_\delta$.  We see that the streaks initially grow faster than free-stream velocity and $A_s/u_0(t)$ increases until about $t=-2$. Subsequently, there is an equilibrium stage until about $t=-0.5$, in which streaks and free-stream velocity grow in proportion, hence $A_s/u_0(t)$ remains approximately constant. Following this phase, the normalized streak amplitudes increase dramatically, as steady rollers keep pumping momentum into streaks, while free-stream velocity  stagnates and decelerates. The critical streak amplitudes calculated by \cite{Vaughan:2011ho} ($A_s^c=0.152$) and by \cite{Andersson:2001dm} ($A_s^c=0.26$) for the initiation of instabilities on steady streaks are also shown in figure~\ref{fig:wMax}b. The discrepancy between these critical values is due to differences in the shapes of streaks employed in these works. It is observed that the case $A=15$ remains below the critical streak amplitudes in the FPG stage and is expected to be stable in this stage. In contrast, cases $A=50$~and~$100$  exceed the critical values already in the FPG stage, hence can develop early instabilities. These observations will be confirmed in \S~\ref{sec:breakdown} using secondary stability analysis. 

 \begin{figure}
\begin{flushright}
\includegraphics{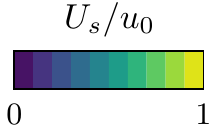}
\includegraphics{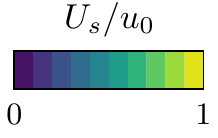}
\includegraphics{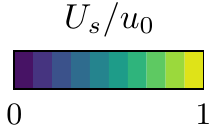}
\includegraphics{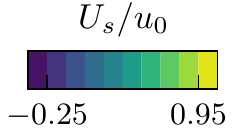}
\includegraphics{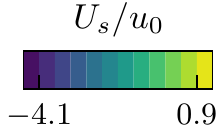}~~~~~~
\end{flushright}
\begin{center}
\includegraphics{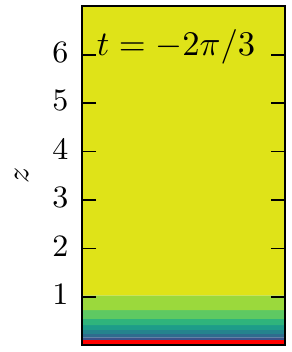}
~\includegraphics{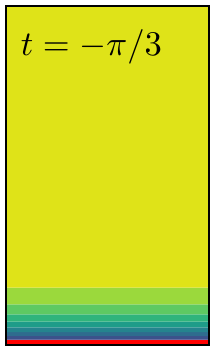}
~\includegraphics{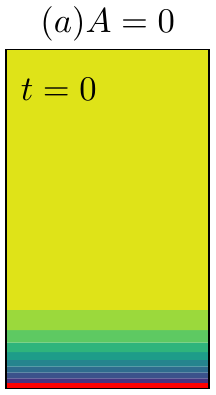}
~\includegraphics{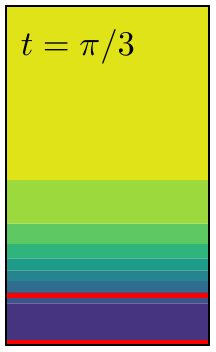}
~\includegraphics{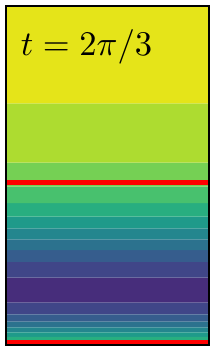}
\vspace{1mm}

\includegraphics{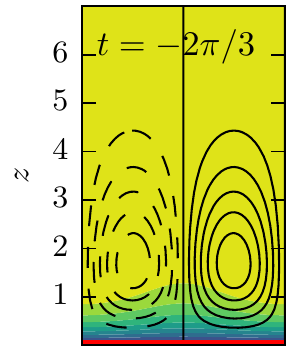}
~\includegraphics{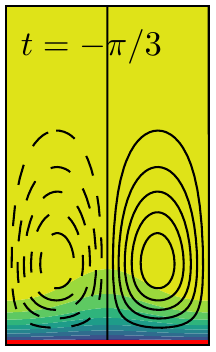}
~\includegraphics{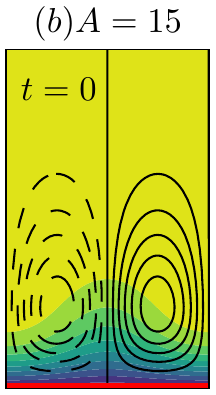}
~\includegraphics{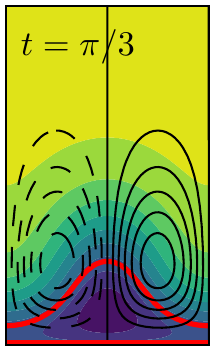}
~\includegraphics{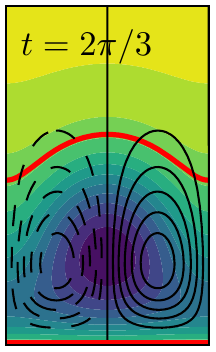}
\vspace{1mm}

\includegraphics{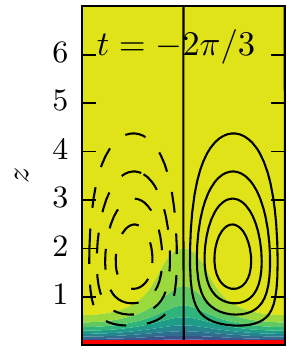}
~\includegraphics{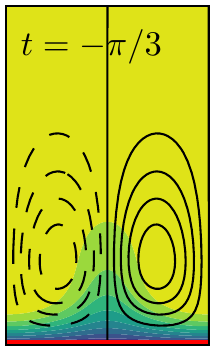}
~\includegraphics{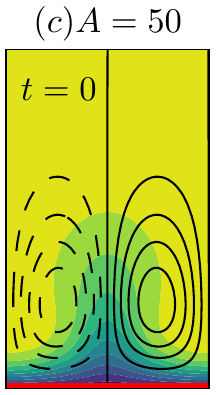}
~\includegraphics{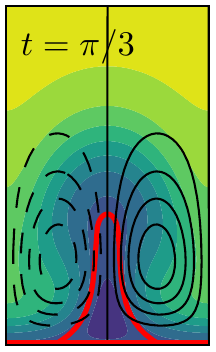}
~\includegraphics{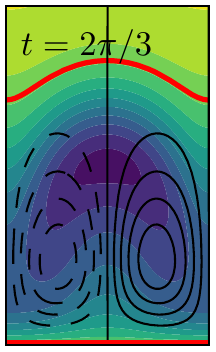}
\vspace{1mm}

\includegraphics{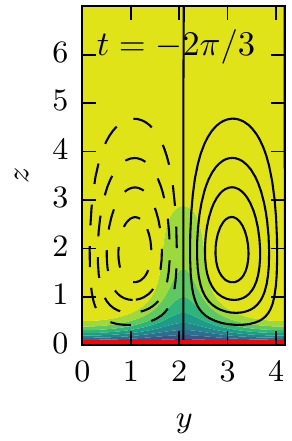}
\includegraphics{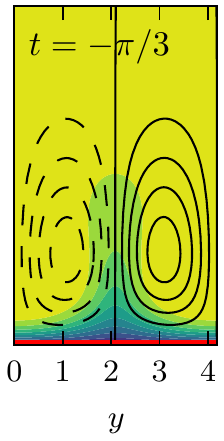}
\includegraphics{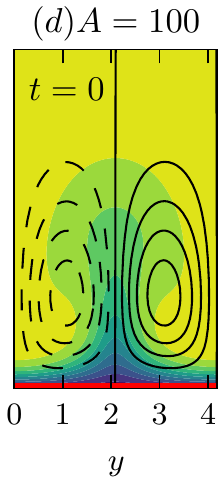}
\includegraphics{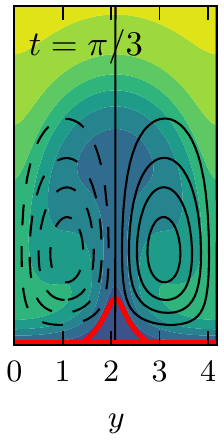}
\includegraphics{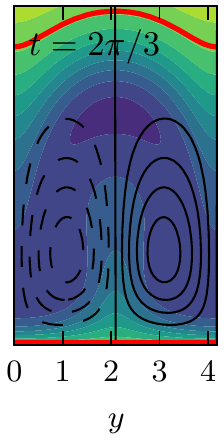}
\end{center}
\caption{\label{fig:nonStreaks15} Nonlinear streaks induced by steady streamwise-constant optimal external excitation $\vec f^{opt}(\alpha=0,\beta=1.5,\omega_f=0,T_f=0)$ with different amplitudes: (a) $A=0$; (b) $A=15$ ($\tilde w_{\max}=2.8/\Rey_\delta$); (c) $A=50$  ($\tilde w_{\max}=9.51/\Rey_\delta$); (d) $A=100$ ($\tilde w_{\max}=18.78/\Rey_\delta$), cf. (\ref{eq:A}). Filled contours show levels of total streamwise velocity scaled with the local phase value of free-stream velocity, i.e., $U_s/u_0(t)$. Each colorbar on top shows the contour levels in the panes below. The thick red contourlines show $u=0$. Black line contours show the streamfunction $\tilde \psi_o$ spanning 9 levels between minimum and maximum values at the phase.} 
\end{figure}

The streaky fields induced by linearly optimal steady streamwise-constant forcing are two-dimensional and have three components, i.e., 
\begin{equation}
\label{eq:UBase}
[U_{s},V_s,W_s](y,z,t):=[U,0,0](z,t)+[\tilde u,0,0](y,z,t)+[0,\tilde v_o,\tilde w_o](y,z),
\end{equation}
where the velocity components associated with the streamwise-constant streak--roller system, $\vec{\tilde u}=[\tilde u,\tilde v_o,\tilde w_o]$,  added to the standard laminar profile ($U$) of the SWBL. The spatial organization of these fields corresponding to the cases $A=15,50$~and~$100$ along with the baseline case ($A=0$) are shown in figure \ref{fig:nonStreaks15}.  Filled contours in figure \ref{fig:nonStreaks15} show the streamwise velocity fields scaled with the local free-stream velocity at the respective phase, $U_s/u_0(t)$, and line contours show levels of $\dbtilde \psi_o$. Increasing nonlinearity with increasing $A$ leads to more uneven streak growth, where low-speed streaks are lifted up to higher fluid layers and narrow down, e.g., compare the figures \ref{fig:nonStreaks15}b,c,d. These elevated low-momentum regions, low-speed streaks, are bounded by internal shear layers with strong local spanwise and vertical variations. 

\begin{figure}
\begin{center}
\includegraphics{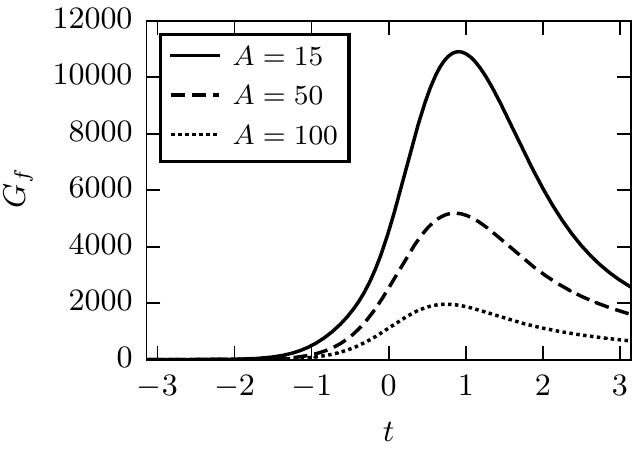}
\end{center}
\caption{ \label{fig:EuE} Nonlinear flow response measured by the gain (\ref{eq:nonLinGain}) for the cases $A=15,50$~and~$100$ at $\Rey_\delta=2000$. The velocity fields for these cases are presented in figures~\ref{fig:nonStreaks15}b-d. }
\end{figure}

In the case of steady streamwise-constant excitation, the gain in the nonlinear regime reads
\begin{equation}
\label{eq:nonLinGain}
	G_f(t;\beta,A,\Rey_\delta,\alpha=0,\omega_f=0)=\frac{E_{\mathcal V}(t)}{E_{\mathcal Vo}},
\end{equation}
 where
\begin{equation}
\label{eq:EintNon}
E_{\mathcal V}(t)=\frac{1}{2}\int_0^\infty\int_{0}^{2\pi/\beta} [\tilde u^2+\tilde v_o^2+\tilde w_o^2]\mathrm dy\mathrm dz,
~~~E_{\mathcal Vo}=\frac{1}{2}\int_0^\infty\int_{0}^{2\pi/\beta}  \left[\tilde v_o^2+\tilde w_o^2\right]\mathrm dy\mathrm dz,
\end{equation}
are the integrated kinetic energy at a time $t$ and the initial energy, respectively.  Although the streak--roller system given by $[\tilde u,\dbtilde v_o,\dbtilde w_o]$ fields is $\Rey_\delta$-independent, the actual flow is $\Rey_\delta$-dependent, as the physical cross-stream components are $[\tilde v_o,\tilde w_o]=A/\Rey_\delta[\dbtilde v,\dbtilde w]$. Therefore, similarity with respect to $A$ only applies to streaks not to rollers, hence the dependency of $G_f$ on Reynolds number. We note that $\tilde v_o^2$ and $\tilde w_o^2$ are two orders lower in Reynolds number than $\tilde u^2$ and therefore, can be neglected in sufficiently high Reynolds numbers, i.e., $E_{\mathcal V,u}\approx E_{\mathcal V}$, where $E_{\mathcal V,u}$ is the integrated streamwise kinetic energy. Consequently, if we define  a gain with similarity variables
\begin{equation}
\label{eq:nonLinGain}
	\tilde G_f(t;\beta,A,\alpha=0,\omega_f=0)=\frac{ E_{\mathcal V,u}(t)}{\tilde E_{\mathcal Vo}},
\end{equation}
where
\begin{equation}
\tilde E_{\mathcal Vo}=\frac{1}{2}\int_0^\infty\int_{0}^{2\pi/\beta}  \left[\dbtilde v_o^2+\dbtilde w_o^2\right]\mathrm dy\mathrm dz,
\end{equation}
then the quadratic dependency of $G_f$ on Reynolds number can be shown explicitly
\begin{equation}
	G_f=\left(\frac{\Rey_\delta}{A}\right)^2\tilde G_f.
\end{equation}
Figure~\ref{fig:EuE} shows the gains for $A=15,50$~and~$100$ at $\Rey_\delta=2000$. The nonlinear saturation greatly limits the amplification of streaks in higher amplitudes, and therefore, the gains are much lower. The streaks amplify until about $t\approx0.9$, and then decay with the flow reversal.


\section{Secondary instability of the nonlinear streaks} \label{sec:transition}
The modulated SWBL featuring the streaks presents a new laminar state, which can be analyzed for its linear stability. In zero-pressure-gradient (ZPG) boundary layers, once the nonlinearity saturates the streaks, their evolution downstream is slow. Therefore, the streamwise velocity on a downstream cross section is assumed steady and streamwise-invariant, and employed as the base state in the secondary stability analysis \citep{Andersson:2001dm}. We examine the stability of nonlinear streaks in SWBL using a similar approach. The main challenge in adapting a secondary stability analysis to the present problem is the transient nature of SWBL -- streaky base states evolve strongly under the effect of strong dynamic and aperiodic pressure gradients. In this regard, a suitable approach to identify temporally unstable regions beneath the wave is the quasi-static stability analysis, in which each instantaneous state is analyzed separately for momentary instabilities  \citep{chen1971stability}. Such an assumption is only valid, if the growth rate of the instability is significantly higher than that of the mean flow. In this regard, the quasi-static assumption is not ideal to draw instability balloons of a transient flow, as close to critical conditions the growth rates become too small to satisfy the quasi-static assumption \citep{kerczek1974linear}. Nevertheless, the approach is quite useful to identify rapidly growing instabilities relevant for transition to turbulence in the SWBL.  The quasi-static assumption for the present flow will be validated in \S~\ref{sec:breakdown} using DNS.

\cite{Blondeaux:2012ei} applied the quasi-static stability analysis to SWBL by considering two-dimensional perturbations growing on one-dimensional, one-component base profiles $U(z,t)$. Here, the analysis is extended to two-dimensional streaky fields with three components $[U_{s},V_s,W_s](y,z,t)$ as shown in (\ref{eq:UBase}). We consider three-dimensional tertiary perturbations of the form
\begin{equation}
\label{eq:fAnsatz}
[ u^\prime, v^\prime, w^\prime, p^\prime](x,y,z,t)=\mathrm {Re}\{[\hat u^\prime,\hat v^\prime,\hat w^\prime,\hat p^\prime] (y,z,t) \mathrm e^{\ii (\alpha x-\int_0^t\omega(\tau) \mathrm d \tau)}\},
\end{equation}
where $\alpha$ are real wavenumbers, and $\omega=\omega_{\mathrm r}+\ii\omega_{\ii}$ are associated complex frequencies. Introducing these perturbations to incompressible Navier--Stokes equations, and linearizing the resulting equations around the two-dimensional frozen base state $[U_{s},V_{s},W_{s}](y,z,t)$, we obtain
\begin{align}
\label{eq:linNSSec1}
-\frac{2}{\Rey_\delta}\ii\omega \hat u^\prime+\ii\alpha U_{s} \hat u^\prime+V_{s}\frac{\partial \hat u^\prime}{\partial y}+W_{s}\frac{\partial \hat u^\prime}{\partial z}+\hat v^\prime\frac{\partial U_{s}}{\partial y}+\hat w^\prime\frac{\partial U_{s}}{\partial z}-\frac{1}{\Rey_\delta}\Delta \hat u^\prime &=-\ii \alpha  \hat p^\prime,   \\
\label{eq:linNSSec2}
-\frac{2}{\Rey_\delta}\ii\omega \hat v^\prime+\ii\alpha U_{s} \hat v^\prime+V_{s}\frac{\partial \hat v^\prime}{\partial y}+W_{s}\frac{\partial \hat v^\prime}{\partial z}+\hat v^\prime\frac{\partial V_{s}}{\partial y}+\hat w^\prime\frac{\partial V_{s}}{\partial z}-\frac{1}{\Rey_\delta}\Delta \hat v^\prime &=-\frac{\partial \hat p^\prime}{\partial y},   \\
\label{eq:linNSSec3}
-\frac{2}{\Rey_\delta}\ii\omega \hat w^\prime+\ii\alpha U_{s} \hat w^\prime+V_{s}\frac{\partial \hat w^\prime}{\partial y}+W_{s}\frac{\partial \hat w^\prime}{\partial z}+\hat v^\prime\frac{\partial W_{s}}{\partial y}+\hat w^\prime\frac{\partial W_{s}}{\partial z}-\frac{1}{\Rey_\delta}\Delta \hat w^\prime &=-\frac{\partial \hat p^\prime}{\partial z},   \\
\label{eq:linNSSec4}
\ii\alpha \hat u^\prime+\frac{\partial \hat v^\prime}{\partial y}+\frac{\partial \hat w^\prime}{\partial z}=0&,
\end{align}
These equations are complemented with boundary conditions $\hat u^\prime=\hat v^\prime=\hat w^\prime=0$ on $z=0$ and $z\rightarrow\infty$.  The system of equations (\ref{eq:linNSSec1})-(\ref{eq:linNSSec4}) can be written shortly as $L^\prime(U(t)) \vec {\hat q}^\prime=\omega(t) \vec {\hat q}^\prime$, where $\vec {\hat q}^\prime(y,z,t)=[\hat u^\prime,\hat v^\prime,\hat w^\prime,\hat p^\prime] (y,z,t) $ .  Using the quasi-static approximation, an eigenvalue problem is defined by freezing the operator $L^\prime$ at a temporal station $t=t_s$ and solving for $\omega(t_s)$ and $\vec {\hat q}^\prime(y,z,t_s)$, the eigenvalue and associated eigenfunction at $t_s$.   The solution of the eigenproblem at each phase is obtained with Nektar++ using an Arnoldi algorithm developed by \cite{tuckerman2000bifurcation} and \cite{barkley2008direct}. The reader is referred to \cite{Rocco:2014tc} for the details.  For several representative cases, we have cross-checked the calculated leading eigenvalues with the ones obtained by simple power iterations, where the linear equations with random initial conditions are integrated in time until convergence to the leading eigenvalue is achieved. Excellent agreements are found for the imaginary parts of eigenvalues (growth rates) but noticeable differences in real parts (frequencies) are observed. Since real parts obtained with Arnoldi method appeared to be more sensitive to configuration parameters, we shall use the results from power iterations in the presentation of phase velocities.   


Symmetries in the two-dimensional streaky fields allow six different groups of secondary perturbations: symmetric or antisymmetric (with respect to base-flow streak)  perturbations associated with fundamental, subharmonic and detuned modes. Fundamental modes share the same spanwise periodicity with the base flow, subharmonic modes have twice the periodicity of the base flow, and detuned modes corresponds to the remaining modes, cf. \cite{reddyJFM98} for mathematical details. In ZPG boundary layers, the most growing eigenvalues are associated with sinuous perturbations. Using an inviscid analysis, \cite{Andersson:2001dm} reported that fundamental and subharmonic sinuous modes have comparable growth rates with subharmonic modes.  The eigenfunctions of both fundamental and subharmonic sinuous modes concentrate in regions on the elevated shear layers around low-speed streaks with very similar patterns, but fundamental modes are slightly more localized (figures 12a and 12b in \cite{Andersson:2001dm}). \cite{Ricco:2011dj} employed a more comprehensive model accounting for the effects of spatial growth and unsteadiness of streaks, and found that fundamental sinuous modes are the most unstable modes.  In stochastic FST-driven bypass-transition scenarios, the transition is usually initiated by breakdown of a single streak \citep{Hack:2014bd}. Therefore, the fundamental instabilities, with their more localized nature around the base streak, are possibly more representative for practical situations.  In this regard, we consider only fundamental-mode instabilities in the present analysis, which allows us to use a periodic domain with a single streak. The spatial discretization on $y-z$ plane is identical with \S~\ref{sec:streaks}. Only leading eigenvalues and eigenmodes are calculated. 

\begin{figure}
\begin{center}
\includegraphics{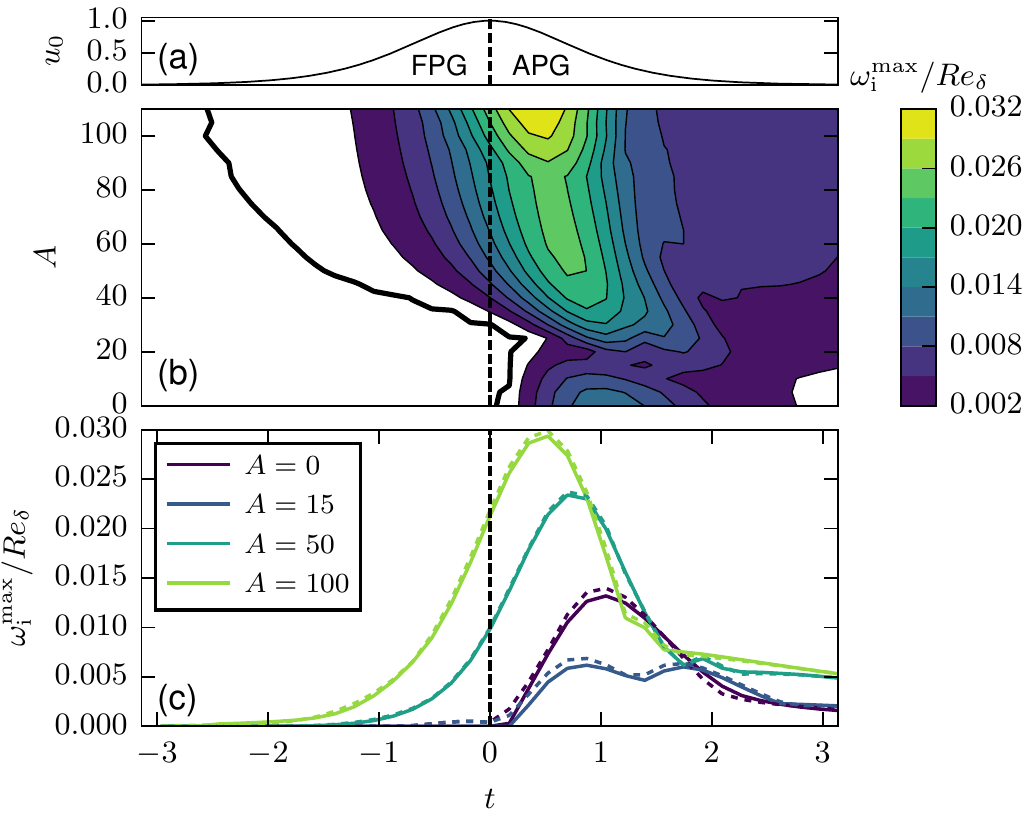}
\end{center}
\caption{\label{fig:staMaps} Stability of the SWBL perturbed by the linearly optimal excitation $f^{opt}(\alpha=0,\beta=1.5, \omega_f=0,T_f=0)$. (a) Free-stream velocity. (b) Contours of leading imaginary eigenvalues at $\Rey_\delta=2000$ calculated with separate stability analysis at each $(A,t)$ using quasi-static assumption. The presented values are the maximum values along all streamwise wavenumbers ($\alpha$). Thick contourlines show $\omega_{\ii}/\Rey_\delta=0.0001$. (c) The maximum growth rates of the nonlinear streaks forced with $A=0,15,50$~and~$100$ (cf. figure~\ref{fig:nonStreaks15}). ($-$): $\Rey_\delta=2000$; ($--$): $\Rey_\delta=4000$. }
\end{figure}

As in \S~\ref{sec:streaks}, we only  consider nonlinear streaks induced by streamwise-constant time-invariant excitation $f^{opt}(\alpha=0,\beta=1.5, \omega_f=0,T_f=0,\Rey_\delta)$. Figure~\ref{fig:staMaps}b shows the maximum leading eigenvalues along all streamwise wavenumbers, $\omega_{\ii}^{\max}(t)=\max_{\alpha}\{\omega_{\ii}(\alpha,t)\}$ for varying excitation amplitudes $A$, (\ref{eq:A}). The  stability boundaries are demonstrated with thick contourlines in figure~\ref{fig:staMaps}b. A slightly positive value $\omega_{\ii}^{\max}/\Rey_\delta=0.0001$ is employed, as the exact neutral curve ($\omega_{\ii}^{\max}=0$) is quite noisy in early times. It should be stressed that the exact location of the neutral curve has little practical bearing, as the quasi-static assumption is only physically relevant when the instabilities evolve faster than the base flow. For weak perturbations in the range $A\lesssim35$, it is observed that the boundary layer remains stable throughout the FPG stage and becomes unstable only when the APG stage starts, i.e., the crest of the wave passes the probing location.  With further increasing forcing amplitude ($A>35$), the instabilities spread also to the FPG stage. The growth rates for the stronger streaks in this range rise until the flow reversal in the early APG stage and peak roughly at a phase when maximum streak amplitudes are observed (cf. figure~\ref{fig:EuE}).  


The maximum growth rates calculated at $\Rey_\delta=2000$ and $\Rey_\delta=4000$ are shown separately in figure~\ref{fig:staMaps}c for the cases $A=0,15,50$~and~$100$, for which the streak amplitudes are plotted in figure~\ref{fig:wMax}b and the base states are presented in figure~\ref{fig:nonStreaks15}. The results for the unperturbed case $A=0$ corresponds to the growth rates of the primary two-dimensional instabilities. These orderly instabilities take place in the APG stage following the emergence of inflectional profiles. The details of the primary instabilities are well documented elsewhere, e.g., \cite{Blondeaux:2012ei} and \cite{Sadek:2015jm}. The case $A=15$ represents a case in the regime with weak streak amplitudes (figure~\ref{fig:wMax}b). Interestingly, the secondary instability in this case has lower growth rates than the baseline primary instability. In the peaking phase ($t\approx 1$), the growth rate in $A=15$ is about the half of the one in $A=0$. These reduced growth rates suggest a damping mechanism introduced to the flow by weak-amplitude streaks. This will be elaborated later.  The last two cases, $A=50$~and~$100$, showcase the results for strong streaks that can develop instabilities already in the FPG stage. As the FPG stage is linearly stable in the unperturbed SWBL, these early instabilities in the FPG stage imply a possibility for a subcritical transition.  The seeding phase of the instabilities in $A=50$ and $A=100$  roughly corresponds to the phases when the streak amplitudes exceeds the critical threshold given by \cite{Vaughan:2011ho}, cf. figure~\ref{fig:wMax}b. We further see in figure~\ref{fig:staMaps}c that the scaled growth rates, $\omega_{\ii}^{\max}/\Rey_\delta$, for the instabilities in $A=50$~and~$100$ are independent of Reynolds number for the selected range (cf. blue and green lines). There is a very weak dependence on Reynolds number for the other cases, cf. the discrepancy at the start of APG stage for magenta and red lines. The reader is referred to \cite{sadek2015mechanisms} for the analysis of Reynolds-number dependency of the primary two-dimensional instabilities ($A=0$).

\begin{figure}
\begin{center}
\includegraphics{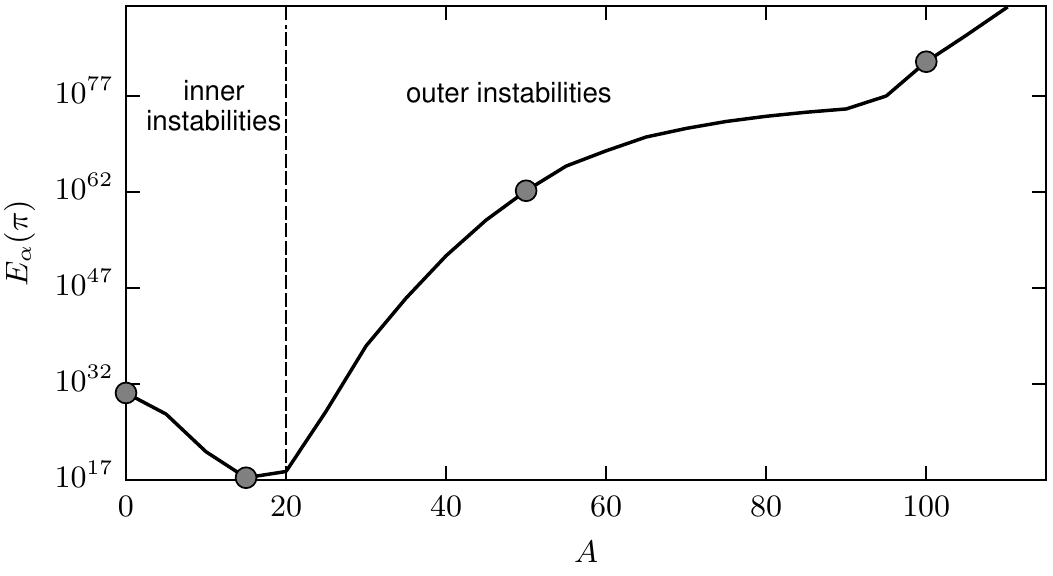}
\end{center}
\caption{\label{fig:EPi} The overall growth measured by the modal perturbation kinetic energy density $E_\alpha$ at $t=\pi$, cf. (\ref{eq:modalE}). The initial energy density is set to $E_0=1$. The most unstable streamwise wavenumber ($\alpha$) is employed at each $A$. The symbols refer to the cases $A=0,15,50$~and~$100$, which are further elaborated in figures~\ref{fig:alphaOmega} and \ref{fig:criticalLayer}.} 
\end{figure}

\begin{figure}
\begin{center}
\includegraphics{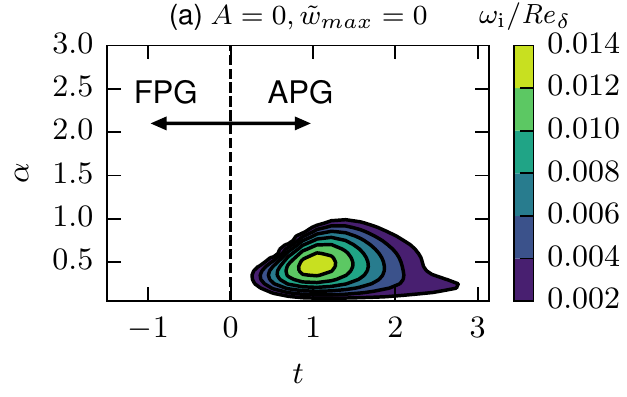}
\includegraphics{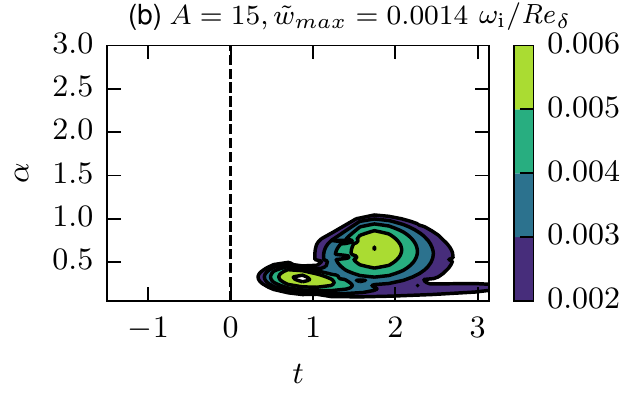}

\includegraphics{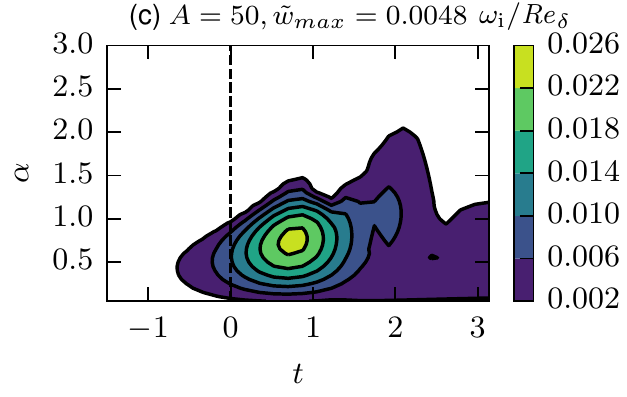}
\includegraphics{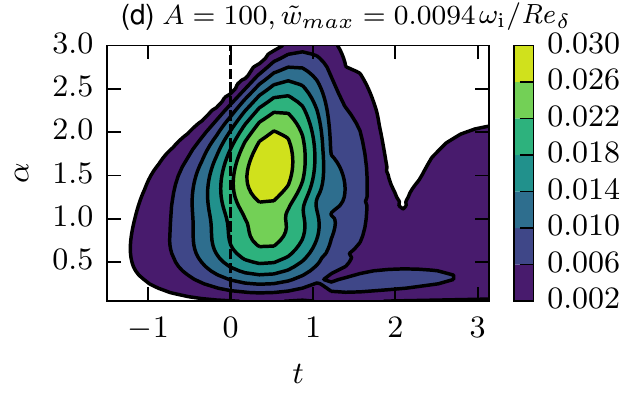}
\end{center}
\caption{\label{fig:alphaOmega} The variation of leading imaginary eigenvalues with respect to streamwise wavenumber for the streaks induced by  linearly optimal excitation $f^{opt}(\alpha=0,\beta=1.5, \omega_f=0,T_f=0,\Rey_\delta=2000)$. (a) Baseline case with $A=0$ (\ref{eq:A}) corresponding to the roller magnitude $\tilde w_{\max}=0$ (\ref{eq:wMax}); (b) $A=15,\tilde w_{\max}=0.0014$; (c) $A=50,\tilde w_{\max}=0.0048$; (d) $A=100,\tilde w_{\max}=0.0094$. Color scales of the contours are different in each pane and are shown in separate colorbars next to the panes.}
\end{figure}

It was observed in figure~\ref{fig:staMaps} that the streaks have a dual role in the transition to turbulence beneath solitary waves: they can be stabilizing or destabilizing depending on their amplitude. This can be elaborated by considering the overall growth of the perturbation energy in time. Using the ansatz (\ref{eq:fAnsatz}), the energy at a mode is expressed by
\begin{equation}
\label{eq:modalE}
E_\alpha(t)=E_0  \mathrm e^{2\int_0^t\omega_{\ii}(\tau) \mathrm d \tau},
\end{equation}
where $E_0$ is the initial energy density at the mode. Figure~\ref{fig:EPi} demonstrates the variation of $E_\alpha$ at $t=\pi$ with respect to $A$. In this figure, the most unstable $\alpha$ for each respective $A$ is evaluated.  Three different instability regimes are observed: (i) primary instability ($A=0$); (ii) inner-instability regime (say $0<A<20$, or $0<\tilde w_{\max}<3.76/\Rey_\delta$ using figure~\ref{fig:wMax}a), and (iii) outer-instability regime ($A>20$, or $\tilde w_{\max}>3.76/\Rey_\delta$). We employed here the naming convention proposed by \cite{Vaughan:2011ho}, where ``inner'' and ``outer'' refer to the location of the critical layer. In the inner-instability regime, streaks are weak but still effective in mixing the momentum of the base profiles and introducing a damping effect. Consequently, there is a reduction in the growth rates of instabilities. A similar stabilizing effect  is observed in flat-plate boundary layers  when moderate-amplitude streaks are superposed on TS waves  \citep{cossu2004tollmien,liu2008floquet}. The temporally unstable phases in the inner-instability regime roughly overlaps with the baseline instabilities in the undisturbed regime (cf.~figure~\ref{fig:staMaps}b), which suggests a modified instability of similar nature, where the primary driving mechanism is vertical shear ($\partial U_s/\partial z$). We will show later the inner instabilities develop in regions nearer to the wall, where the vertical shear is strong, hence the name ``inner''.   In the third regime, the streaks are strong enough to develop secondary shear layer instabilities in the elevated outer zones. The overall growth due to these instabilities rise dramatically between $20<A<60$. Afterwards, there is a saturation range until $A\approx90$, in which increased forcing does not lead to substantial growth. However, with further increasing $A$ there is another receptive regime for $A>90$.  

\begin{figure}
\begin{center}
\includegraphics{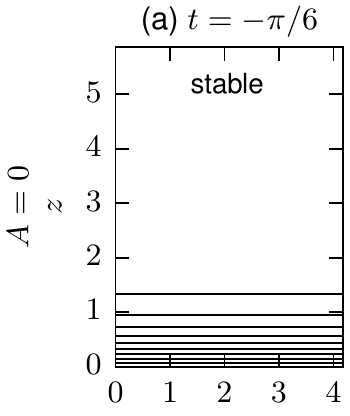}
\includegraphics{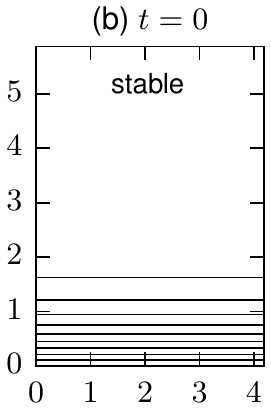}
\includegraphics{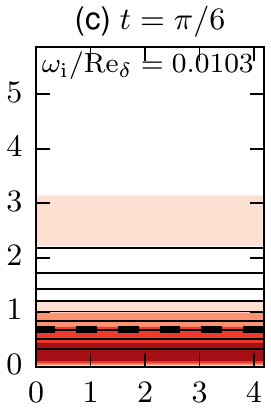}
\includegraphics{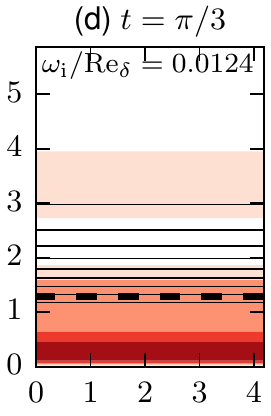}

\includegraphics{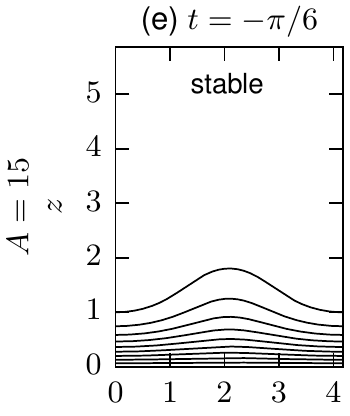}
\includegraphics{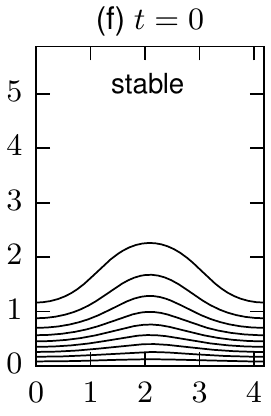}
\includegraphics{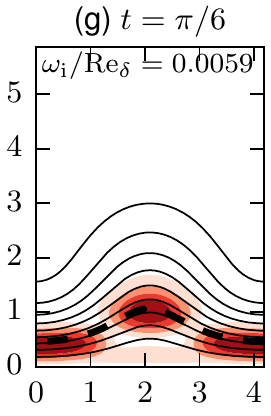}
\includegraphics{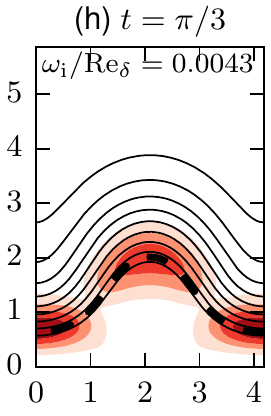}

\includegraphics{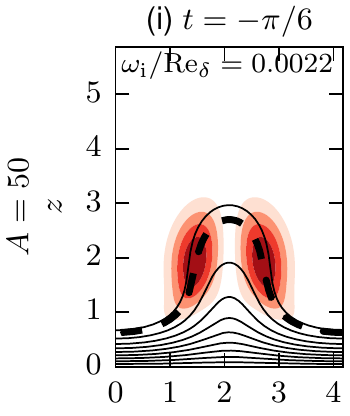}
\includegraphics{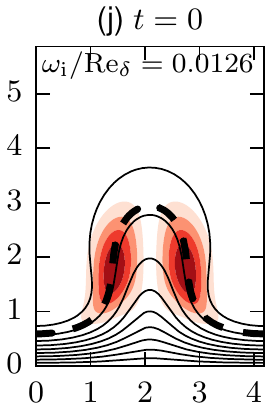}
\includegraphics{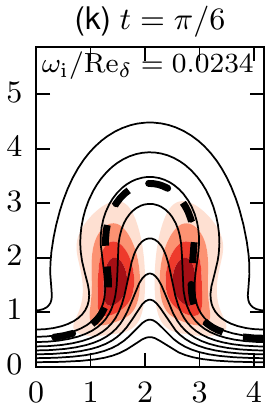}
\includegraphics{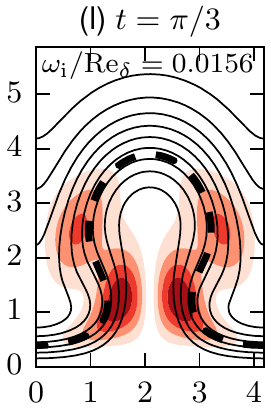}

\includegraphics{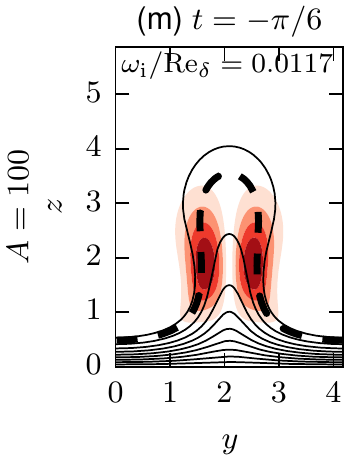}
\includegraphics{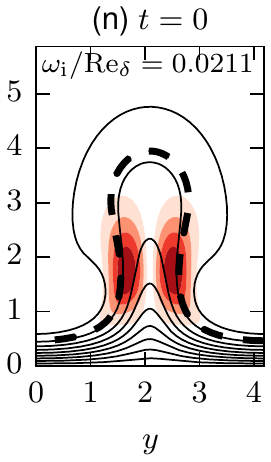}
\includegraphics{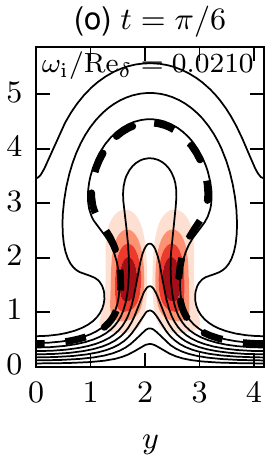}
\includegraphics{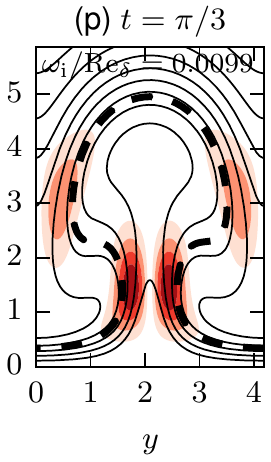}

\end{center}
\caption{\label{fig:criticalLayer} Filled red contours show four different levels of the modulus of the streamwise component of the leading eigenmodes, $|\hat { u}^\prime|$, calculated at four different times $t=-\pi/6,0,\pi/6,\pi/3$. The base states are nonlinear streaks induced by the linearly optimal excitation $f^{opt}(\alpha=0,\beta=1.5, \omega_f=0,T_f=0,\Rey_\delta=2000)$ for excitation magnitudes (a-d): $A=0,\alpha=0.45$; (e-h): $A=15,\alpha=0.35$; (i-l): $A=50,\alpha=0.75$; (m-p): $A=100,\alpha=0.75$. Contour lines show  the streamwise component of the base flow scaled with the free-stream velocity at the respective phase, $U_s(y,z,\theta_s=-40^\circ)/u_0(t)=\{0.1:0.1:0.9\}$. The dashed lines represent the critical velocity level $U_s^c$, at which $U_s=c_r$. }
\end{figure}

 Figure~\ref{fig:alphaOmega} demonstrates the variation of growth rates with respect to streamwise wavenumbers $\alpha$ for the cases $A=0,15,50$~and~$100$ calculated at $\Rey_\delta=2000$. It is observed that the primary instabilities have relatively longer wavelength  peaking at wavenumbers around $\alpha\approx0.45$ (figure~\ref{fig:alphaOmega}a), where the outer instabilities concentrate in shorter wavelengths, e.g., $\alpha\approx\beta/2\approx0.75$ for $A=50$ (figure~\ref{fig:alphaOmega}c) and $\alpha\approx\beta\approx1.5$ for $A=100$ (figure~\ref{fig:alphaOmega}d). We see that the case $A=100$ is sensitive to a wider range of instabilities, e.g., $A=50$ is mostly stable for the short wave perturbations with $\alpha>1.5$ whereas $A=100$ is still very unstable at this range. It is this additional support for highly unstable short-wavelength modes, which gives rise to a second receptive regime for $A>90$ in figure~\ref{fig:EPi}.  The case $A=15$, belonging to inner-instability regime, shows a mixed behavior. Right after the flow reversal, there is a peak at $t\approx1$ and $\alpha\approx0.35$ but then the growth of the long waves stagnate (figure~\ref{fig:alphaOmega}b). Meanwhile, we observe another growth region concentrated at shorter wavelengths close to $\alpha\approx0.75$. This is the typical wavenumber for the short wave outer instabilities, which suggests that outer instabilities are  influential at mid to late APG stage in the case $A=15$. However, as the overall growth ($E_\alpha$) at $\alpha=0.35$ is higher, the inner-instabilities are the dominant secondary mechanism at $A=15$.

\begin{figure}
\begin{center}
\includegraphics{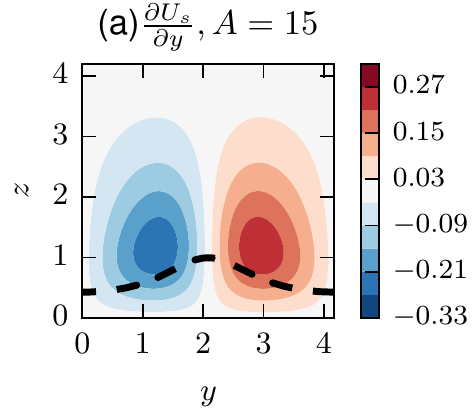}
~~~\includegraphics{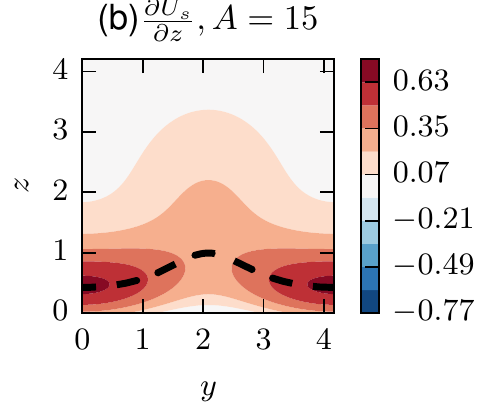}

\includegraphics{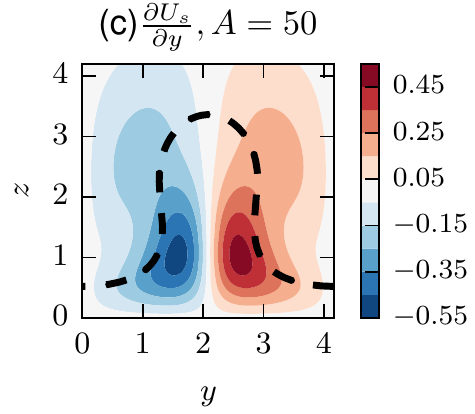}
~~~\includegraphics{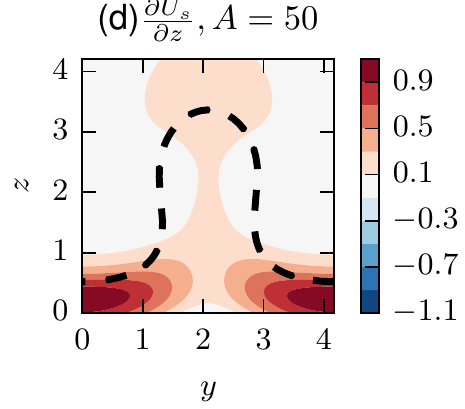}
\end{center}
\caption{ \label{fig:shearLayers} Derivative  fields of the base streamwise velocity $U_s$ at $t=\pi/6$. (a) $\partial U_s/\partial y$ in $A=15,\alpha=0.35$, (b) $\partial U_s/\partial z$ in $A=15,\alpha=0.35$, (c) $\partial U_s/\partial y$ in $A=50,\alpha=0.75$, (d) $\partial U_s/\partial z$ in $A=50,\alpha=0.75$. The dashed lines show the critical layers, where $U_s=c_r$. The streaks are induced by steady streamwise-constant optimal external excitation $\vec f^{opt}(\alpha=0,\beta=1.5,\omega_f=0,T_f=0,\Rey_\delta=2000)$.}
\end{figure}

We now turn to the the nature of instabilities, e.g., the symmetry patterns, phase velocities, amplification mechanisms. To this end, we select the cases with maximum growth rates in each instability regime, i.e., $(A=0,\alpha=0.45)$;  $(A=15,\alpha=0.35)$; $(A=50,\alpha=0.75)$; $(A=100,\alpha=0.75)$. Figure~\ref{fig:criticalLayer} demonstrates the spatial distribution of eigenmodes using the modulus of streamwise components. It is observed that the unstable modes in primary-instability ($A=0$) and inner-instability ($A=15$) regimes extend to the whole spanwise extent of the periodic domain, cf. figures~\ref{fig:criticalLayer}c,d,g,h. In contrast, the eigenmodes are located around the elevated low-speed streaks and are of more localized nature in the outer-instabilities regime, cf. figures~\ref{fig:criticalLayer}i-p. The instabilities in streaky flows are generated by inviscid mechanisms due to inflection points in the shear layers. In these instabilities, critical layers, where $U_s^c:U_s=c_r=\omega_{\mathrm r}/\alpha$, form. The eigenmodes concentrate in the critical layers, where they convect with local mean velocity $U_s^c$, cf. dashed lines in figure~\ref{fig:criticalLayer}. Streaky boundary layers consist of spanwise and vertical shear layers associated with $\partial U_s/\partial y$ and $\partial U_s/\partial z$, respectively. These are shown in figure~\ref{fig:shearLayers} for two representative cases  $(A=15,\alpha=0.35)$ and $(A=50,\alpha=0.75)$ at time $t=\pi/6$. We see that the critical layer in the inner-instability regime develops on the vertical shear layer close to the wall (figure~\ref{fig:shearLayers}b), while the critical layer in the outer-instability regime is located in the elevated spanwise shear layers around the low-speed streak (figure~\ref{fig:shearLayers}c).

The spanwise and vertical shear layers in streaky boundary layers have a dampening effect, when the principle instability does not develop on them. This can be understood by breaking down the generation of the modal perturbation kinetic energy into individual components. Following \cite{cossu2004tollmien}, we express the globally integrated budget as follows
\begin{equation}
\frac{\partial  e_{\mathcal V}}{\partial t}= p_{\mathcal V,y}+ p_{\mathcal V,z}- d_{\mathcal V},
\end{equation}
where $e_{\mathcal V}$ is the total perturbation kinetic energy, $p_{\mathcal V,y}$ is the total production rate due to spanwise shear, $p_{\mathcal V,w}$ is the total production rate due to vertical shear and $d_{\mathcal V}$ is the total dissipation rate. Here we neglect the contributions related to base spanwise ($V_s$) and vertical ($W_s$) velocities as $\tilde v_o$ and $\tilde w_o$ are an order of magnitude smaller in $\Rey_\delta$. If we consider the perturbations associated with a single instability mode, we can express individual terms by  $[e_{\mathcal V}, p_{\mathcal V,y}, p_{\mathcal V,z}, d_{\mathcal V}]=[\hat e_{\mathcal V}, \hat  p_{\mathcal V,y}, \hat p_{\mathcal V,z}, \hat d_{\mathcal V}]\ee^{2\omega_{\ii}t}$, where 
\begin{equation}
[\hat e_{\mathcal V}, \hat  p_{\mathcal V,y}, \hat p_{\mathcal V,z}, \hat d_{\mathcal V}]=\frac{1}{\lambda_y}\int _0^{\lambda_y}\int_0^{\infty} [\hat {E},\mathcal{\hat{P}}_{v},\mathcal {\hat{P}}_{w}, \hat{\mathcal D}] \mathrm dy\mathrm dz,
\end{equation}
$\lambda_y=2\pi/\beta$ and 
\begin{align}
\label{eq:tkeTerms}
\begin{split}
	\hat E=\frac{2}{\Rey_\delta}(\hat u^{\prime*}\hat u^\prime+\hat v^{\prime*}\hat v^\prime+\hat w^{\prime*}\hat w^\prime), &~~~~~ \mathcal {\hat D}=\frac{2}{\Rey_\delta}(\hat \varepsilon^{\prime*}\hat \varepsilon^\prime+\hat \zeta^{\prime*}\hat \zeta^\prime+\hat \eta^{\prime*}\hat \eta^\prime),\\
	\hat P_y= -(\hat u^{\prime*}\hat v^\prime+\hat v^{\prime*}\hat u^\prime)\frac{\partial U_s}{\partial y}, &~~~~~~~\hat P_z= -(\hat u^{\prime*}\hat w^\prime+\hat w^{\prime*}\hat u^\prime)\frac{\partial U_s}{\partial z}.
\end{split}
\end{align}
In the dissipation term $\hat {\mathcal D}$,  the components of the perturbation vorticity $[\varepsilon^\prime,\zeta^\prime,\eta^\prime]$ are employed. Now the growth rate of the instability can be expressed as
\begin{equation}
\label{eq:tkeOmega}
\omega_\ii=\frac{1}{2 \hat e_{\mathcal V} }(\hat  p_{\mathcal V,y}+ \hat p_{\mathcal V,z}- \hat d_{\mathcal V}).	
\end{equation}

\begin{figure}
\begin{center}
\includegraphics{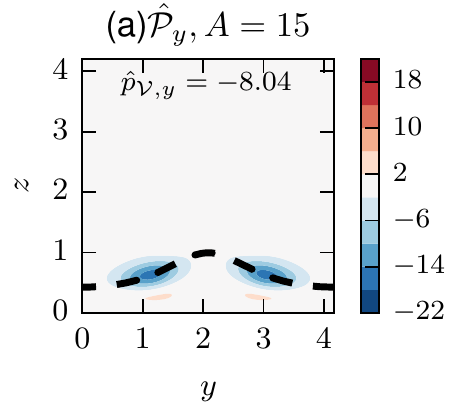}
\includegraphics{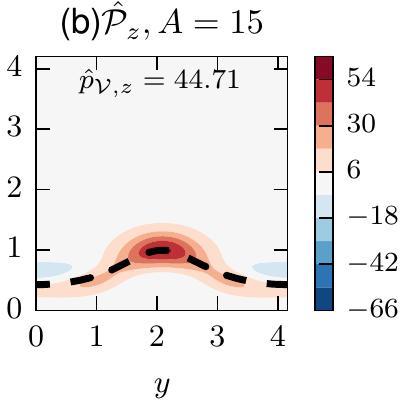}
\includegraphics{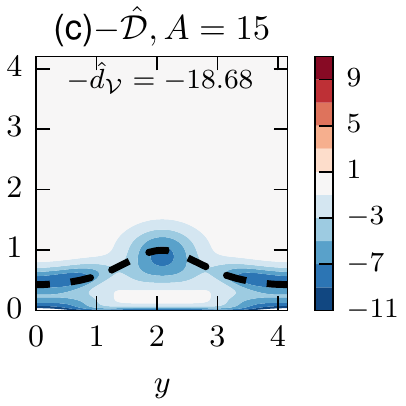}

\includegraphics{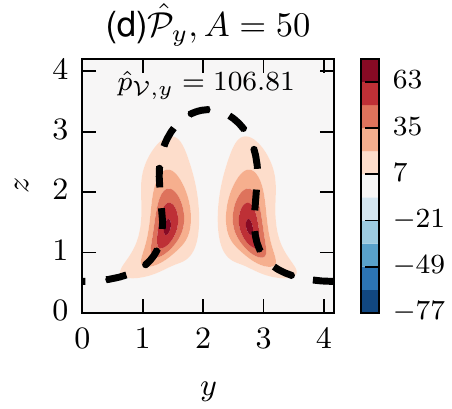}
\includegraphics{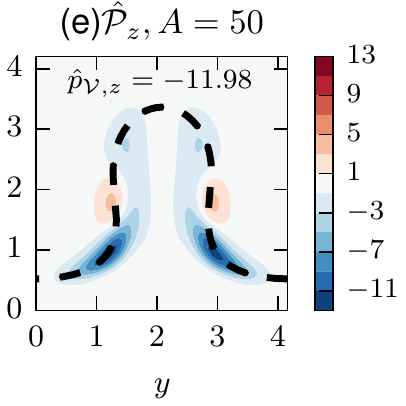}
\includegraphics{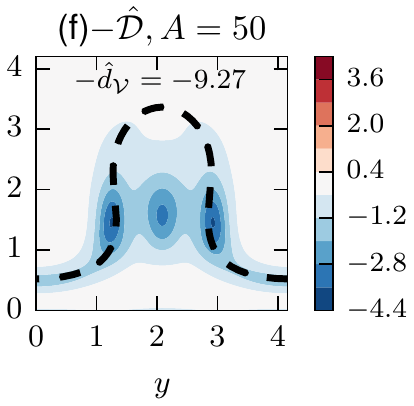}
\end{center}
\caption{ \label{fig:tke} Production and dissipation rates of perturbation kinetic energy (cf. \ref{eq:tkeTerms}) for cases $A=15,\alpha=0.35$ (a-c) and $A=50,\alpha=0.75$ (d-f) at time $t=\pi/6$, cf. figures~\ref{fig:criticalLayer}g,k. The energy of the perturbations is normalized to unity, i.e., $e_{\mathcal V}=1$, in the calculation of fields. The streaky base states are induced by steady streamwise-constant optimal external excitation $\vec f^{opt}(\alpha=0,\beta=1.5,\omega_f=0,T_f=0,\Rey_\delta=2000)$. (a,d): Production rate due to spanwise shear ($\hat {\mathcal P}_y$); (b,e): Production rate due to vertical shear ($\hat {\mathcal P}_z$); (c,f): Dissipation rate ($\hat {\mathcal D}$). The integrated contributions of the fields, $\hat p_{\mathcal V,y},\hat p_{\mathcal V,y}$ and $\hat d_{\mathcal V}$,  are also presented in the respective panels. }
\end{figure}

Figure~\ref{fig:tke} shows the production and dissipation fields for the cases  $A=15,\alpha=0.35$ and $A=50,\alpha=0.75$ at time $t=\pi/6$. The total integrated values, $\hat p_{\mathcal V,y},\hat p_{\mathcal V,y},\hat d_{\mathcal V}$,  are also presented in the respective panels of the fields. The energy of the perturbations is normalized to unity, i.e., $e_{\mathcal V}=1$. In the case of inner instability ($A=15$), the production due to vertical shear feeds the growth (figure~\ref{fig:tke}b), while the production due to spanwise shear has negative contributions to the total budget, i.e., has a stabilizing effect (figure~\ref{fig:tke}c).  Vice versa is true for the outer instability -- the spanwise shear drives the instability, while vertical shear trying to counteract it, cf.  figures~\ref{fig:tke}d,e. The degree of dualism between the two shear production mechanisms and the dissipation rate determines together the growth rate of the instability, cf. (\ref{eq:tkeOmega}). In case $A=15$, the shear-damping effect is stronger with $\left | \hat p_{\mathcal V,y}\right |$ being about $20\%$ of  $p_{\mathcal V,y}$. Furthermore, the dissipation rate is also higher in this case, as the perturbations are closer to the wall where viscous effects are more pronounced.

The counteracting role of vertical and spanwise shear layers in boundary layers has been well documented in steady flows. \cite{reddyJFM98} discussed the stabilizing effect of the vertical shear on the outer instabilities developing on high-amplitude streaks. The same shear damping applies in outer-instability regime in SWBLs (figures~\ref{fig:tke}d,e).  For low-amplitude streaks,  \cite{cossu2004tollmien} has reported inner TS-like instabilities  with reduced growth rates. They suggested the negative contributions from $\hat p_{\mathcal V,y}$ as the primary mechanism behind the stabilizing effect of low-amplitude streaks. This term vanishes in the unperturbed boundary layer ($\partial U/\partial y=0$), while the vertical production rates remain at similar magnitude, hence the higher growth rate of the undisturbed TS instability. The same stabilization mechanism is also effective in the inner-instability regime of SWBLs. However, we note that another mechanism contributing to the reduction of growth rates is the increase in dissipation due to three dimensionality. Orderly instability modes are two-dimensional and have one component vorticity ($\zeta^\prime$), which yields lower dissipation rates, cf. $\hat{\mathcal D}$ in (\ref{eq:tkeTerms}). 

\begin{figure}
\begin{center}
\includegraphics{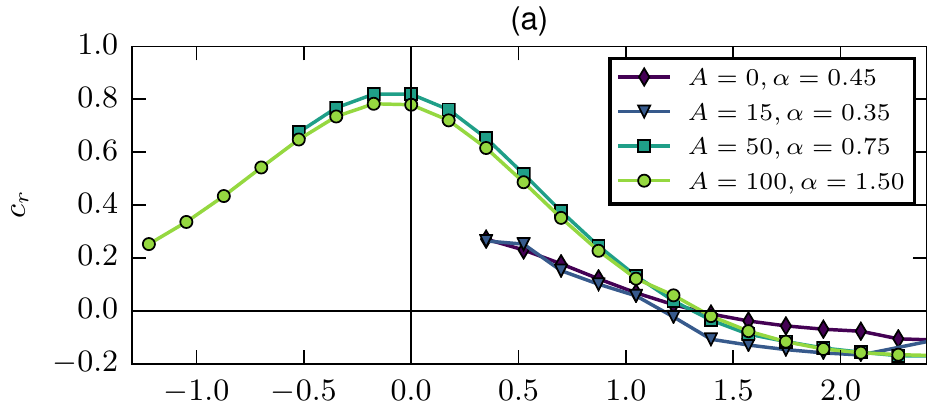}
\includegraphics{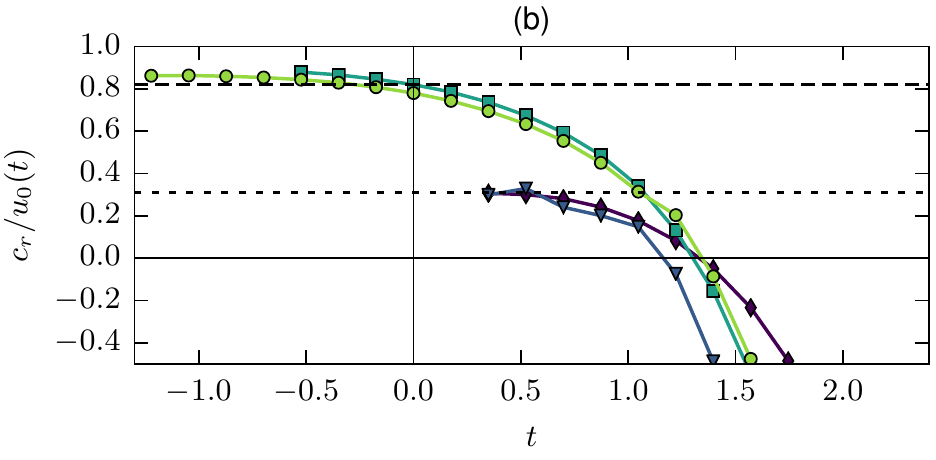}
\end{center}
\caption{\label{fig:cr} Phase velocities $c_r=\omega_{\ii}/\alpha$ for the cases presented in figure~\ref{fig:criticalLayer}. (a) Absolute values. (b) Normalized values with the local free-stream velocity at the respective phases. Dashed line shows $c_r/u_0(t)=0.82$, which is the phase velocity calculated by \cite{Andersson:2001dm} for a outer streak instability in a ZPG boundary layer. Dotted line shows $c_r/u_0(t)=0.31$, which is the phase velocity calculated by \cite{cossu2004tollmien} for  an inner (modified TS-like) instability.}
\end{figure}

\begin{figure}
\begin{center}
\includegraphics[scale=0.175]{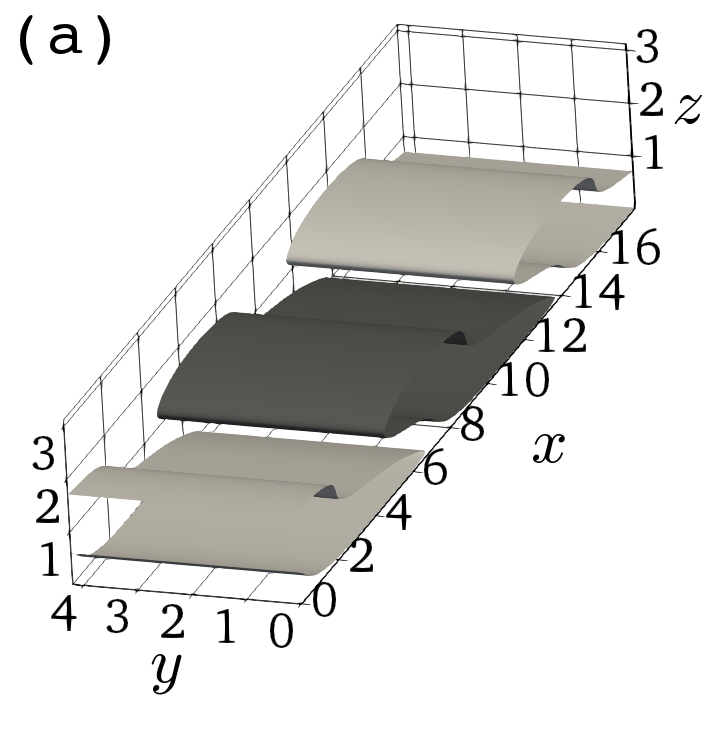}
\includegraphics[scale=0.175]{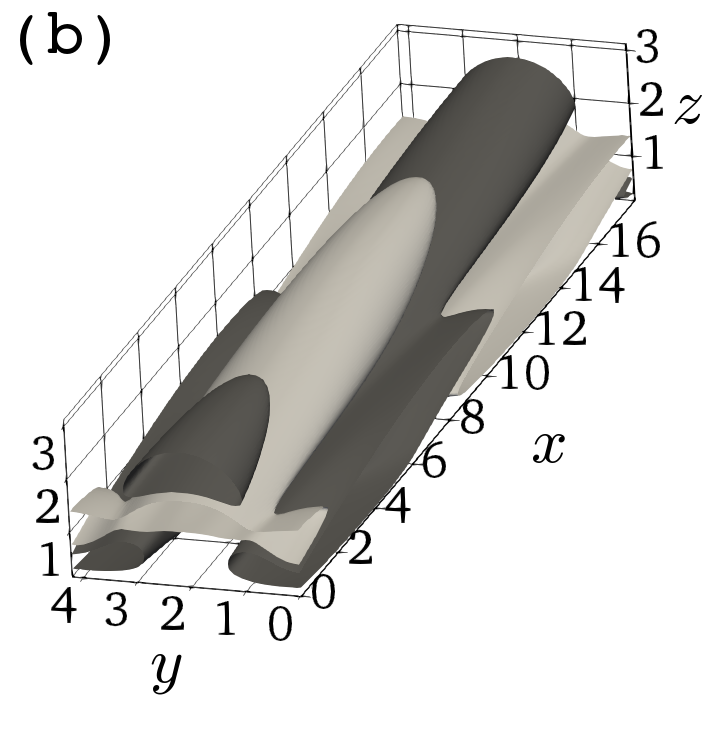}
\includegraphics[scale=0.175]{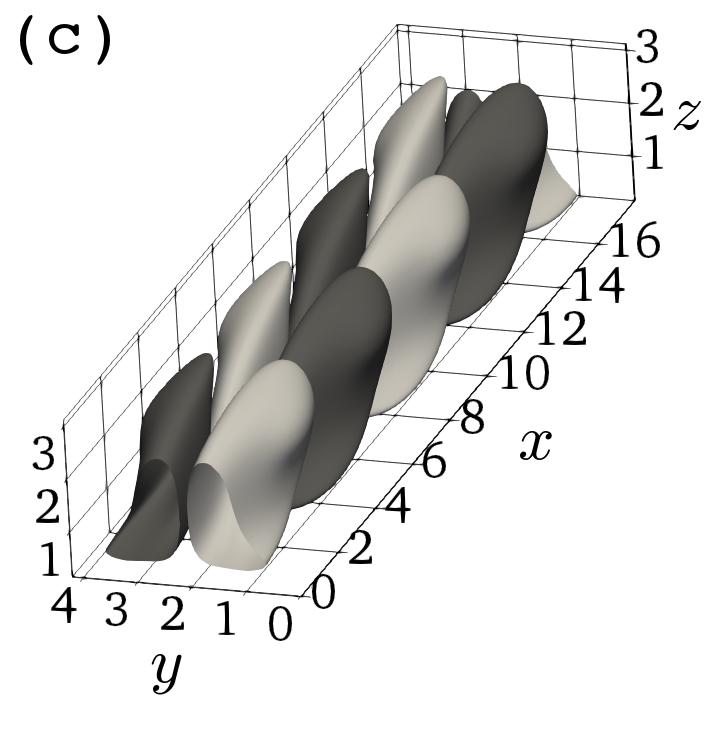}
\caption{\label{fig:eigenmodes}  Symmetry patterns of instabilities are shown using the isosurfaces of the streamwise component of eigenmodes. (a) two-dimensional mode at streamwise wavenumber $\alpha=0.45$ growing on baseline flow with $A=0$ at time $t=\pi/3$ (cf. figure~\ref{fig:criticalLayer}d); (b) varicose mode at $\alpha=0.35$   growing on low-amplitude streaks with $A=15$ at $t=\pi/3$ (cf. figure~\ref{fig:criticalLayer}h);  (c) sinuous mode at $\alpha=0.75$ growing on streaks with $A=50$ at $t=\pi/6$ (cf. figure~\ref{fig:criticalLayer}k). (light): negative isosurface; (dark): positive isosurface. }
\end{center}
\end{figure}

The characteristics of the instabilities can be further elaborated by studying the phase velocities and symmetry patterns. Figure~\ref{fig:cr} plots the phase velocities for the cases presented in figure~\ref{fig:criticalLayer}, where figure~\ref{fig:cr}a is scaled with $u_{0,m}^*$ and figure~\ref{fig:cr}b is in local scaling with the free-stream velocity at the respective phase ($c_r/u_0(t)$). Additionally, the phase velocity of the outer streak instability in \cite{Andersson:2001dm} ($c_r=0.82$) and the inner TS-like instability in \cite{cossu2004tollmien} ($c_r=0.31$) are also shown in figure~\ref{fig:cr}b for reference. We observe in figure~\ref{fig:cr}b that the phase velocity of outer instabilities in FPG stage has very close values to their counterparts in ZPG boundary layers, cf. light-colored lines in $t<0$ in figure~\ref{fig:cr}b. The deceleration in the APG stage has a dramatic effect on the phase velocity of these modes, and they decay rapidly in the APG stage. The instabilities generated by the vertical shear ($A=0,15$) follows initially the phase velocity calculated by \cite{cossu2004tollmien} but then decays also rapidly with flow reversal in the vicinity of the wall. 

Figure~\ref{fig:eigenmodes} shows the symmetry pattern of the primary, inner and outer instabilities using the streamwise velocity of instances shown in figures~\ref{fig:criticalLayer}d,h,k. The primary instability is as expected two dimensional (cf. figure~\ref{fig:eigenmodes}a). On the other hand, the inner instability is a varicose mode strongly tilted in the streamwise direction (cf. figure~\ref{fig:eigenmodes}b). A varicose symmetry is also reported in \cite{cossu2004tollmien}) for the inner TS-like instability. Finally, we see in figure~\ref{fig:eigenmodes}c that the outer instability is in the form of sinuous mode. Sinuous modes are the most unstable mode in ZPG boundary layers \citep{Andersson:2001dm,Ricco:2011dj} and also commonly observed as breaking mode of streaks in experiments (e.g. \cite{Mans:2007gw}).

\section{Direct numerical simulations} \label{sec:breakdown}
 In \S~\ref{sec:streaks}, we analyzed the dynamics of streamwise-constant streaks  in a two-dimensional domain ($\vec x=[x=0,y,z]$).  In this section, we investigate the response of the streaks to small-amplitude background noise to validate the quasi-static assumption employed in \S~\ref{sec:transition} and to investigate breakdown stage in transition. To this end, three-dimensional perturbations are introduced and direct numerical simulations are conducted. The optimal forcing configuration remain identical to the one employed in \S~\ref{sec:streaks} and \S~\ref{sec:transition}, i.e., $f^{opt}(\alpha=0,\beta=1.5, \omega_f=0,T_f=0,\Rey_\delta)$. 

\begin{table}
\begin{center}
\begin{tabular}{ c c  c c c c c c c }  
\hline
Case &$A$&$A_r$&$\Rey_\delta$& $T_r$& $\alpha_0$&$\beta_0$&$L_x\times L_y\times L_z$ &$N_x\times N_y\times N_z$   \\
\hline
A0c1 &0&$10^{-1}$&4000&0&$0.225$& - &$2\pi/\alpha_0\times 4\pi/\alpha_0\times 20$  & $400\times 320\times 480$  \\
A0c2 &0&$10^{-6}$&4000&0&$0.225$& - &$2\pi/\alpha_0\times 4\pi/\alpha_0\times 20$  & $400\times 320\times 480$  \\
A0c3 &0&$10^{-12}$&4000&0&$0.225$& - &$2\pi/\alpha_0\times 4\pi/\alpha_0\times 20$  & $400\times 320\times 480$  \\
A15c1 &15&$10^{-1}$&4000&$0$&$0.175$& 1.5 &$2\pi/\alpha_0\times 2\pi/\beta_0\times 20$  & $400\times 320\times 480$  \\
A15c2 &15&$10^{-6}$&4000&$0$&$0.175$& 1.5 &$2\pi/\alpha_0\times 2\pi/\beta_0\times 20$  & $400\times 320\times 480$  \\
A15c3 &15&$10^{-12}$&4000&$0$&$0.175$& 1.5 &$2\pi/\alpha_0\times 2\pi/\beta_0\times 20$  & $400\times 320\times 480$  \\
A50c1 &50&$10^{-6}$&2000&$-\pi$&$0.375$& 1.5 &$2\pi/\alpha_0\times 2\pi/\beta_0\times 20$  & $160\times 320\times 480$  \\
A50c2 &50&$10^{-12}$&2000&$-\pi$&$0.375$& 1.5 &$2\pi/\alpha_0\times 2\pi/\beta_0\times 20$  & $160\times 320\times 480$  \\
A100c1 &100&$10^{-6}$&2000&$-\pi$&$0.375$& 1.5 &$2\pi/\alpha_0\times 2\pi/\beta_0\times 20$  & $160\times 320\times 480$  \\
A100c2 &100&$10^{-9}$&2000&$-\pi$&$0.375$& 1.5 &$2\pi/\alpha_0\times 2\pi/\beta_0\times 20$  & $160\times 320\times 480$  \\
A100c3 &100&$10^{-11}$&3000&$-\pi$&$0.375$& 1.5 &$2\pi/\alpha_0\times 2\pi/\beta_0\times 20$  & $160\times 320\times 480$  \\

 \end{tabular}  
\end{center} 
\caption{Computational details of simulations. $\beta_0$ is the spanwise wavenumber of the excitation. $A_r$ is the amplitude of random tertiary perturbations seeded at time $T_r$.  Each element has in total $(P+1)^2$ degrees of freedom (DOF), where $P=7$ is the order of the Lagrange polynomials. The total number of DOFs in $y$ and $z$ directions is calculated as, e.g. for $y$ direction, $N_{y}=N_{ey}\times (P+1)$, where $N_{ey}$ is the number of elements in the $y$ direction.  In the streamwise direction, $N_x/2$ Fourier modes are employed yielding $N_x$ grid points.  
 } 
\label{tab:cases}
\end{table}

The computational details of the cases are presented in Table~\ref{tab:cases}. We consider four representative forcing amplitudes $A=0,15,50$~and~$100$ as in previous sections and perturb the resulting streaky fields with small-amplitude random noise of amplitude of $A_r$. These random perturbations are seeded at the end of FPG stage $T_r=0$ for the  cases with $A=0$ and $A=15$, as these cases are stable in the FPG stage. For the rest, the tertiary random perturbations are seeded at $T_r=-\pi$. The numerical method employed in \S~\ref{sec:streaks} is extended to three dimensions using a mixed spatial discretization in Nektar++. In this method, a bi-dimensional spectral-element method, previously introduced in \S~\ref{sec:streaks}, is employed in streamwise-wall normal ($y-z$) plane, and global Fourier expansions are considered in the streamwise ($x$) direction \citep{karniadakis1990spectral,bolis2013fourier}. To avoid instabilities due to aliasing errors, the method developed by \cite{Kirby:2006hr} is employed for polynomial expansions, and the $2/3$ rule is employed for the Fourier expansions \citep{boyd2001chebyshev}. The computational domain is a rectangular box with dimensions $[0,L_x]\times[0,L_y]\times[0,L_z]$. Periodic boundary conditions are employed in the streamwise and spanwise directions. The domain contains a single streak in the forced cases. The streamwise extent of the domains is selected to allow growth in the most unstable streamwise wavenumbers. No-slip boundary condition is applied at the bottom wall, and free-slip boundary condition is applied at the top wall. \cite{Verschaeve:2014gh} remarked that a very fine grid resolution is necessary to capture the natural development of two-dimensional instabilities.  Therefore, a finer structured grid than the one in \S~\ref{sec:streaks} is employed to resolve instabilities and turbulence.  The grid densities are everywhere considerably higher than the previous DNS works on SWBL \citep{Vittori:2008gv,OZDEMIR:2013bu}. 

\begin{figure}
\begin{center}
\includegraphics{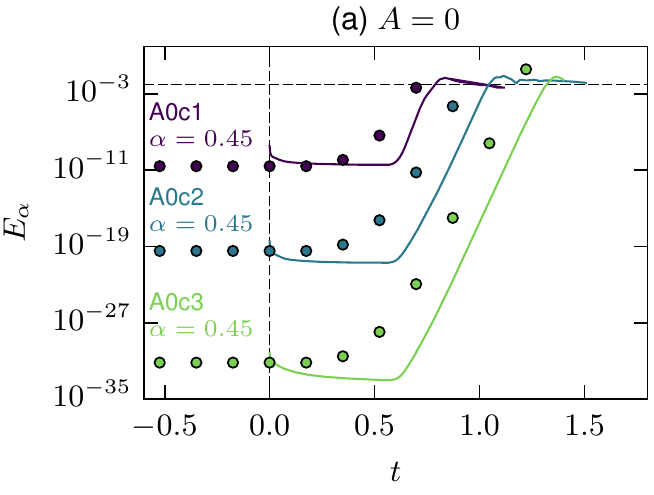}
\includegraphics{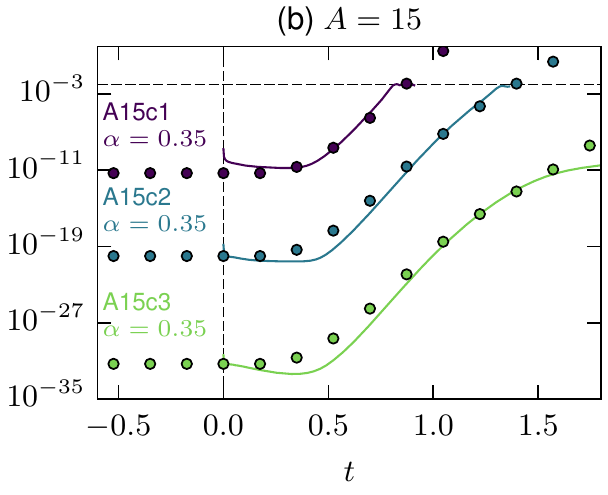}
\includegraphics{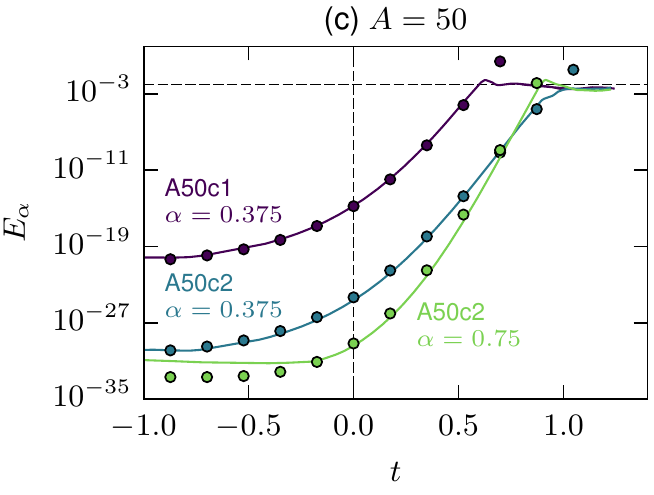}
\includegraphics{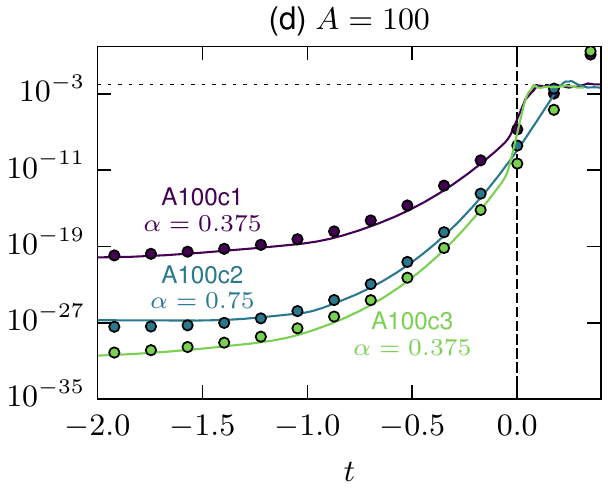}
\end{center}
\caption{\label{fig:dnsEnergy} Growth and nonlinear saturation of secondary instabilities. Lines show the modal energy extracted from DNS, whereas symbols show the ones calculated using the leading eigenvalues of secondary stability analysis (\S~\ref{sec:transition}). (a) Cases A0c1, A0c2, A0c3; (b) Cases A15c1, A15c2, A15c3; (c) Cases A50c1, A50c2; (d) Cases A100c1, A100c2, A100c3, cf. Table~\ref{tab:cases} for case definitions. The horizontal line shows $E_\alpha=10^{-2}$. }
\end{figure}

The energy density in each streamwise mode $\alpha$ is calculated by Fourier transforming the velocity fields in the streamwise direction and integrating the respective energy in the Fourier mode $\vec {\hat u}(\alpha,y,z,t)$ over the domain and then normalizing it, i.e.,
\begin{equation}
\label{eq:modalE2}
E_\alpha(t)=\frac{1}{2L_yL_z}\int_0^{L_z}\int_{0}^{L_y}  \vec {\hat u}^*(\alpha,y,z,t)\cdot \vec {\hat u}(\alpha,y,z,t)\mathrm d y\mathrm dz.
\end{equation}
Since the introduced random perturbations are of small magnitude, we expect linear mechanisms will drive the initial growth of secondary instabilities. Therefore, the modal kinetic energies extracted from the direct numerical simulations should match the ones calculated with the secondary stability analysis with (\ref{eq:EintNon}), if the quasi-static assumption is valid. This is tested in figure~\ref{fig:dnsEnergy} for all cases, where the initial modal energy level ($E_0$) of the linear growth results is adjusted to match the DNS values. An interesting first observation is that the energy growth in all considered cases saturates at about $E^c:=E_\alpha=10^{-2}$ regardless of Reynolds number, saturation phase and the type of instability. This energy level can be considered as the critical threshold for the onset of breakdown. The cases, which cannot reach this level during the wave event, can be assumed still laminar. We observe a good match between DNS and linear stability theory (LST) in the cases with outer instabilities (figure~\ref{fig:dnsEnergy}c,d). In these cases, the long wave instability at $\alpha=0.375$ develops first thanks to its higher growth rate in the FPG stage. Depending on the initial noise amplitude, these long-wave modes can reach the critical level and become the mode of breakdown as in the Case~A50c1, or are overtaken by the shorter wave instability  at $\alpha=0.75$ as in the Case~A50c2, cf. figure~\ref{fig:dnsEnergy}c.

There is also good agreement between DNS and LST in the cases containing inner-instabilities (cf. figure~\ref{fig:dnsEnergy}b). However,  the DNS data stagnates in Case~A15c3 in the interval $t>1.5$ and does not follow the growth dictated by LST anymore (only until $t=1.9$ shown in the figure). This deviation suggests that the instabilities introduce non-negligible deformations to the slow base flow in the late APG stage. Thus, the  quasi-static assumption appears to be inapplicable to later phases of the wave event. The biggest discrepancy between DNS and LST is observed in two-dimensional  baseline instabilities, cf. figure~\ref{fig:dnsEnergy}a. In these cases,  the instabilities in DNS develop with some delay compared to the theoretical predictions.  The stabilizing effect of weak streaks can be clearly seen in figures~\ref{fig:dnsEnergy}a,b.  The onset of transition is substantially delayed in cases with $A=15$ compared to those with $A=0$. In fact, in the cases with lowest initial noise, Case~A15c3 remains laminar, whereas Case~A0c3 breaks into turbulence at about $t\approx1.3$.

\begin{figure}
\begin{center}
\includegraphics{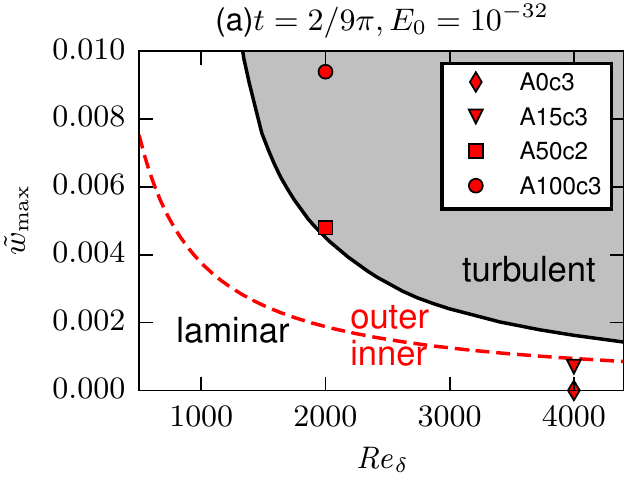}
~~~~~\includegraphics{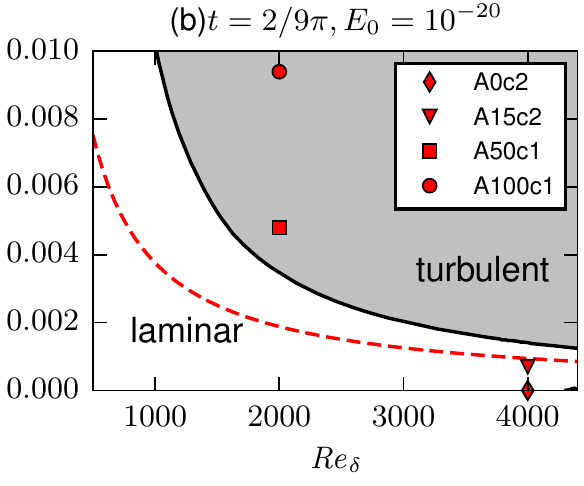}
\includegraphics{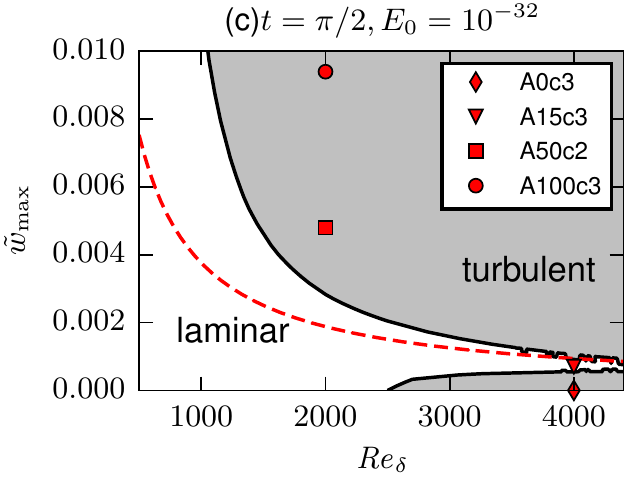}
~~~~~\includegraphics{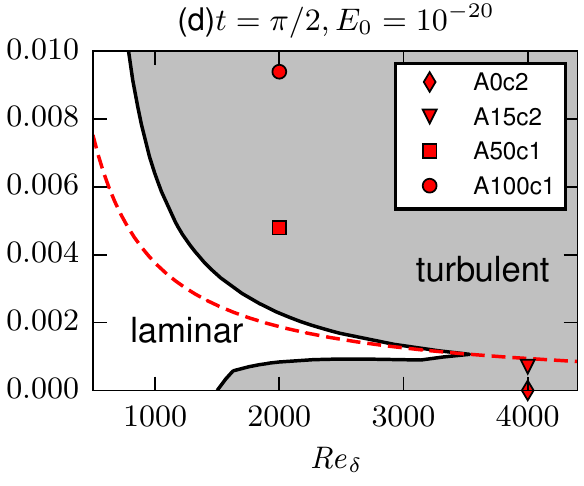}
\end{center}
\caption{\label{fig:threshold} State of the SWBL with respect to Reynolds number ($\Rey_\delta$) and amplitude of steady-roller perturbations ($\tilde w_{\max}$) are shown at two representative times in the APG stage for two different initial tertiary perturbations ($E_0$). The rollers are induced by steady streamwise-constant forcing $f^{opt}(\alpha=0,\beta=1.5, \omega_f=0,T_f=0)$. (a) $t=2/9\pi$, $E_0=10^{-32}$; (b) $t=2/9\pi$, $E_0=10^{-20}$; (c) $t=\pi/2$, $E_0=10^{-32}$;  (d) $t=\pi/2$, $E_0=10^{-20}$. The red dashed lines demonstrate the boundary ($A=20$) between inner and outer instabilities, which corresponds to $\tilde w_{\max}=3.76/\Rey_\delta$ (figure~\ref{fig:wMax}a).} 
\end{figure}

 If we assume that LST results are always applicable and all instabilities are of inviscid nature with constant $\omega_\ii/\Rey_\delta$,  then we can utilize the empirical threshold $E^c$ and the growth rates $\omega_\ii/\Rey_\delta$ calculated at a specific Reynolds number (e.g. $\Rey_\delta=2000$) to extrapolate our results to a wider range of Reynolds numbers and perturbation levels. 
Using this extrapolation, we can generate state diagrams showing whether the flow is laminar or turbulent at an instant $t$. To this end, the state of the flow is a function of four parameters,  $t$, $\tilde w_{\max}$, $\Rey_\delta$, and the initial perturbation energy in the instability mode, $E_0$.  Figures~\ref{fig:threshold} show the flow states with respect to $\tilde w_{\max}$ and $\Rey_\delta$ at $t=2/9\pi$  and $t=\pi/2$ for initial perturbation levels of $E_0=10^{-20}$ and $E_0=10^{-32}$. The cases sharing the same initial perturbation levels are also demonstrated with symbols in the respective diagrams. The boundary between inner and outer instabilities, i.e., $A=20$ corresponding to $\tilde w_{\max}=3.76/\Rey_\delta$ (figure~\ref{fig:wMax}a), is also plotted in figures. As shown in figures~\ref{fig:threshold}a,b the primary and inner streak instabilities are not effective yet at $t=2/9\pi$. At this earlier phase, the transition occurs only due to outer instabilities, which develop when rollers exceed a certain threshold depending on Reynolds number. This is the manifestation of bypass transition. The primary instability modes are bypassed by an early subcritical transition mechanism that is dependent on the magnitude of environment perturbations, $\tilde w_{\max}$ in our model. The flow states are also somewhat sensitive to the amplitude of initial tertiary perturbations ($E_0$) especially in lower Reynolds number, e.g., compare the range $1000<\Rey_\delta<1500$ in figures~\ref{fig:threshold}a,b.  

 When the wave propagates further, the primary and inner instabilities become active, cf. figures~\ref{fig:threshold}c,d. We see that the laminar region protrudes into the turbulent region in the range $\tilde w_{\max}\approx 0.001-0.002$, i.e., the flow remains laminar until relatively high Reynolds numbers in this range. This is the manifestation of the stabilization introduced by weak streaks. For instance, Case~A1c3, which has a roller magnitude of $\tilde w_{\max}=2.8/\Rey_\delta$, remains in the protruded laminar region for $E_0=10^{-32}$, cf. figure \ref{fig:threshold}c. We further observe in figures~\ref{fig:threshold}c,d that the primary and inner instabilities are more sensitive to $E_0$ compared with outer instabilities. These results show that transition to turbulence in the SWBL depends on the amplitude of environment perturbations even in the case of orderly transition with two-dimensional instability modes. Flow-state classifications that are based merely on Reynolds number, e.g., the state charts in \cite{Sumer:2010ce} and \cite{OZDEMIR:2013bu}, have to be extended to include some measure for environment perturbations. 

\begin{figure}
\begin{center}
\includegraphics[scale=0.105]{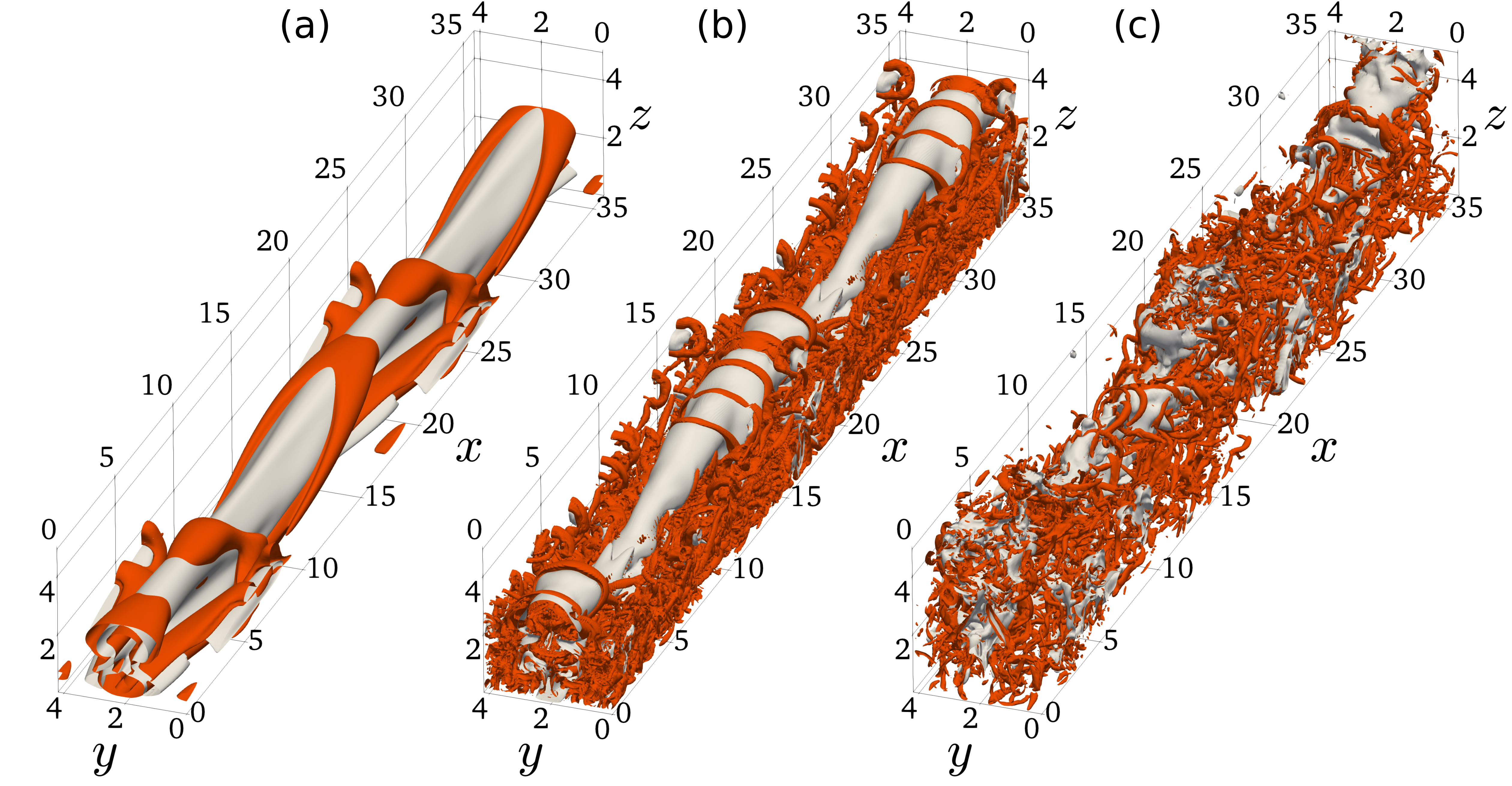}
\caption{\label{fig:breakdown15} Streak breakdown and onset of turbulence in Case~A15c1. White surfaces show the low-speed streak using an isosurface of streamwise fluctuation velocity $\tilde u^\prime=u-\langle u \rangle$, where $\langle u \rangle (z,t)$ is the average value on a plane at $z$.   Colored isosurfaces show the vortical regions using Q criterion. (a) $t=23/90\pi$, $\tilde u^\prime=-0.18$; $Q=0.003$ (b) $t=24/90\pi$, $\tilde u^\prime=-0.18$; $Q=0.15$ ; (c) $t=25/90\pi$, $\tilde u^\prime=-0.18$; $Q=0.55$. }
\end{center}
\end{figure}

\begin{figure}
\begin{center}
\includegraphics[scale=0.06]{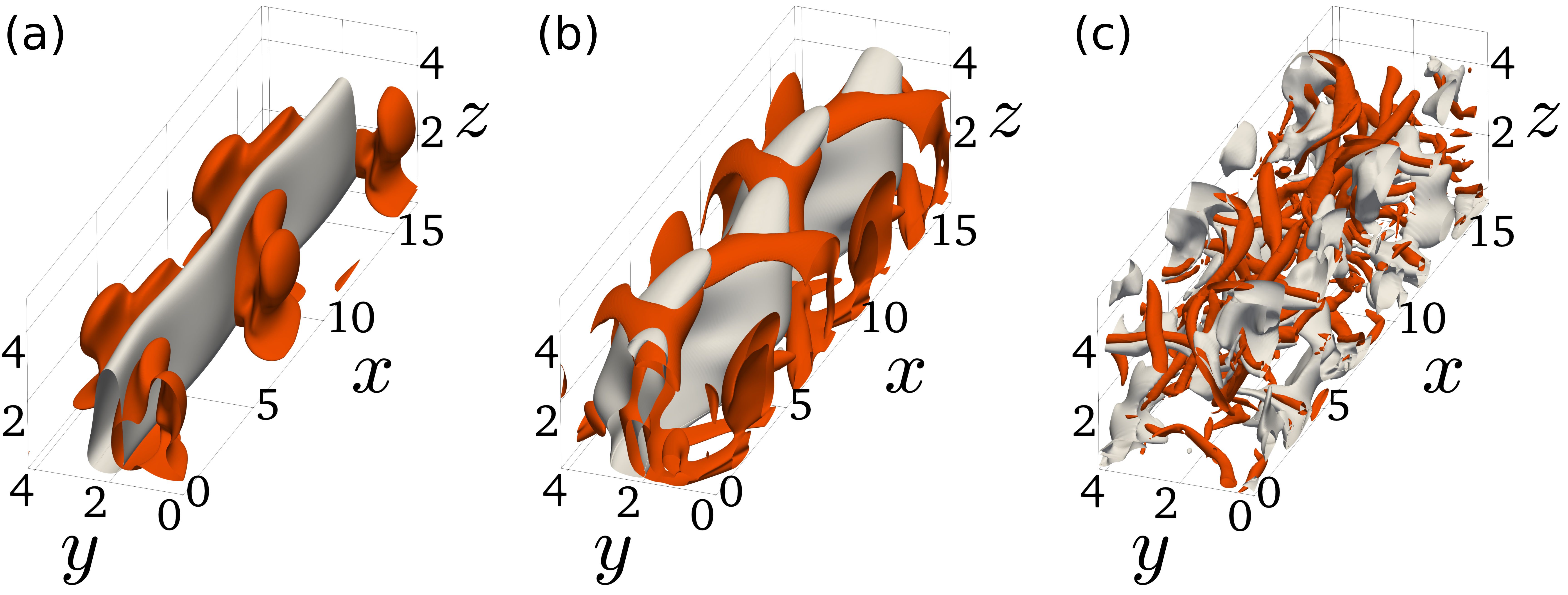}
\caption{\label{fig:breakdown50} Streak breakdown and onset of turbulence in Case~A50c2, cf. figure~\ref{fig:breakdown15} for the definition of surfaces. (a) $t=50/180\pi$, $\tilde u^\prime=-0.26$; $Q=0.006$ (b) $t=52/180\pi$, $\tilde u^\prime=-0.16$; $Q=0.027$ ; (c) $t=55/180\pi$, $\tilde u^\prime=-0.13$; $Q=0.44$.} 
\end{center}
\end{figure}

Figure~\ref{fig:breakdown15} shows the breakdown of inner instability in Case~A15c1. At $t=24/90\pi$ in figure~\ref{fig:breakdown15}a, we see at the center a low-speed streak making  undulations in the downstream direction with wavenumber $\alpha=0.35$. Since the inner instability is of varicose nature, the undulations are symmetric with respect to the streak. Vortical structures around the low-speed streak are also shown in the figure using a positive isosurface of Q-criterion~\citep{Hunt88}. Among several vortical features, $\Lambda$-like vortices can bee seen to accompany the undulating streak, cf. e.g., the region $10<x<20$ in figure~\ref{fig:breakdown15}a. These features are reminiscent of $\Lambda$ vortices developing on streak-modulated instability waves in ZPG boundary layers \citep{liu2008boundary}. Later at $t=25/90\pi$, the breakdown to small scales is initiated in the near-wall layers, while the low-speed streak remains still stable and coherent, cf. small-scale vortices in figure~\ref{fig:breakdown15}b.  Subsequently, chaotic small-scale motions quickly spread everywhere, the streak is disintegrated and the onset of turbulence is completed at $t=26/90\pi$, cf. figure~\ref{fig:breakdown15}c.

The transition to turbulence in Case~C50c2 is demonstrated in figure~\ref{fig:breakdown50}.  Initially at $t=50/180\pi$, we see a low-speed streak at the center of domain occupying the whole streamwise extent, cf. figure~\ref{fig:breakdown50}a. This streak is unstable and exhibits sinuous undulations with a streamwise wavelength corresponding to the dominant outer instability mode at $\alpha=0.75$.  Subsequently, at $t=52/180\pi$, the waviness of streaks is increased and some more tertiary vortical features have emerged, cf. figure~\ref{fig:breakdown50}b. Both vortex and velocity structures appear to be large-scale organized features, thus the flow is still at a laminar transitional state at this phase. Finally, at $t=55/180\pi$, turbulence sets in and chaotic motions are to be seen everywhere in the domain, cf figure~\ref{fig:breakdown50}c. In contrast to Case~A15c1, in which breakdown to small scales is initiated in the inner layers adjacent to stable streaks, the main mechanism of breakdown in Case~C50c2 is the disintegration of the meandering streak in the outer layer.

\section{Conclusions and outlook}\label{sec:conclusion}
We have investigated the transition to turbulence in the bottom boundary layer beneath a solitary wave by means of a simple parallel model taking into account finite amplitude perturbations. The study consists of two steps addressing the receptivity and breakdown stages of transition. In the receptivity step, the most "dangerous" disturbances to which the boundary layer shows the strongest response are found using a linear input-output framework. In this framework, the perturbations are modelled as deterministic body forces. The focus is in particular on early times prior to the flow reversal. The optimal excitation per energy input was found to concentrate on cross-stream components, which are arranged as  streamwise-constant counter-rotating rotational cells. These cells can be either steady or oscillate at frequencies close to the effective wave frequency. This optimally-arranged transverse forces introduce counter-rotating rollers that mix the streamwise momentum of the flow and introduce energetic streamwise-constant streaks via the lift-up effect. We have  then selected a representative case with steady streamwise-constant  configuration at a spanwise wavenumber ($\beta=1.5$) to seed small-amplitude rollers into nonlinear equations. As in the linear case, the dynamics of the rollers are completely decoupled from the base flow and the wave, hence they remain steady throughout the event. Optimally-arranged steady rollers were found to amplify the energy of the streaks with a factor proportional to $\Rey_\delta^2$.  Increasing the amplitude of rollers $\tilde w_{max}$ (\ref{eq:wMax}) leads to streaks with increased asymmetry, where low-speed streaks become narrower and elevate into higher flow regions. 
  
    In the analysis of the breakdown step, we have first investigated the linear secondary stability of perturbed boundary layers to identify the unstable regions beneath the wave. To this end, we employed a quasi-static assumption, which allows a separate stability analysis at each phase using the frozen base flow. Two different streak instabilities were observed, which we denoted as ``inner'' and ``outer'' instabilities after the location of their respective critical layers, a naming convention suggested by \cite{Vaughan:2011ho} for flat-plate boundary layers. The inner instabilities have varicose symmetry and are fed on the vertical shear, thus they have critical layers near to the wall. They are activated in the APG stage at the same phases with the two-dimensional instabilities of the baseline unperturbed flow. Compared to the baseline instabilities, the inner instabilities  have reduced growth rates due to negative production driven by spanwise shear and enhanced dissipation in two-dimensional mode shapes.  The inner instabilities are therefore stabilizing and can delay the transition to turbulence or completely suppress it.   The damping effect is strongest in streaks generated by rollers with magnitude $\tilde w_{\max}\approx2.8/\Rey_\delta$.  In contrast to inner instabilities, outer instabilities were found to be very unstable. They are of sinuous nature and develop around the lifted low-speed streaks in the outer region. These instabilities are driven by the spanwise shear of the base flow. Therefore, they are only active when the low-speed streaks are significantly elevated, which is achieved when the amplitude of the streaks $A_s$ (\ref{eq:As}) exceeds 15\% of the local free-stream velocity at the phase. This can occur already in the FPG stage if the roller-perturbations are strong. Therefore, outer instabilities can lead to a subcritical bypass transition at this stage. The bifurcation point from inner to outer instabilities depends on the roller magnitude and Reynolds number, and is found to be at $\tilde w_{\max}\approx3.8/\Rey_\delta$.  
    
    In the final step of our analysis, the results of secondary stability analysis were verified by means of DNS. We have observed a specific energy level above which breakdown to turbulence occurred in all considered cases. Using this empirical threshold, flow-state diagrams were generated. At a particular phase, the state of the flow, i.e., laminar or turbulent, depends on Reynolds number ($\Rey_\delta$), the roller amplitude ($\tilde w_{max}$) and the initial amplitude of the tertiary perturbation in the secondary instability mode. The state diagrams showed the damping effect of streaks more clearly, e.g., the laminar zone protrudes deep into the turbulent zone for moderate-amplitude perturbations.  For instance, for the case $\tilde w_{\max}=2.8/\Rey_\delta$, the damping mechanism can keep the flow laminar up to very high Reynolds numbers such as $\Rey_\delta=4000$. These observations suggest that the classification of flow states should at least include an additional measure for environment perturbations.  Previous Reynolds-number based classifications are not satisfactory.

We have investigated the effect of finite amplitude perturbations on the transition of a SWBL using an idealized deterministic model, which allows generation of streaks in a controlled setting. A possible future direction is extending the work to a more natural configuration, in which the ambient turbulence and its penetration into boundary layer are considered. In this model, streamwise vortices and streaks will evolve in a stochastic setting. Depending on streak amplitudes, four possible transition scenarios are anticipated: (i) orderly transition when streaks have negligible influence; (ii) delayed transition under low- to moderate-amplitude ambient turbulence, where inner instabilities on moderate-amplitude streaks dominate the APG stage; (iii) bypass-transition under high-amplitude ambient turbulence, where outer instabilities broke streaks into turbulent spots, which then grow, merge and occupy the whole boundary layer; (iv) mixed transition, where any of the prior transition mechanisms can occur at different parts of the boundary layer. The mixed transition can occur  in particular when the amalgamation timescale of turbulent spots is slow. In this case, other transition mechanisms can take place  in laminar regions surrounding spots, e.g., turbulent spots and orderly spanwise rollers coexisted in the APG stage in \cite{Sumer:2010ce}.  Only after full assessment of the amalgamation timescale, it will be clear under which circumstances a complete bypass transition can take place in a SWBL.

\section*{Acknowledgements}
The research reported here has been supported by a grant from Ministry of Education of Singapore to National University of Singapore. The computational work for this article was fully performed on resources of the National Supercomputing Centre, Singapore (https://www.nscc.sg). We thank Mengqi Zhang for providing the starting point of our linear adjoint code and for useful comments. 


\appendix

\section{Derivation of the optimal forcing}\label{app:f}
For our time-dependent problem, the adjoint approach can be utilized using the formal Lagrange method \citep{corbettJFM01,fredi10}. First,  the inner products are defined as
\begin{equation}
 \langle \vec a,\vec b \rangle_{\Omega}=\frac{1}{2}\intz\vec (\vec a^*\cdot \vec b )\mathrm dz +c.c.;  ~~~\langle \vec a,\vec b \rangle_{\overline \Omega}=\frac{1}{2}\intt\intz\vec (\vec a^*\cdot \vec b )\mathrm dz \mathrm dt +c.c.,
\end{equation}
where asterisk denotes complex-conjugated fields and c.c. stands for the complex conjugate of the previous terms in the expression. Subsequently, we associate the following Lagrangian functional to the problem 
\begin{align}
\mathcal L(\vec {\hat q},\vec {\hat q}^+,\vec {\hat f},\sigma)&:=\frac{E(\vec {\hat q}(T_f))}{E(\vec {\hat q}(T_i))}+\langle \vec {\hat q}^{+}, L(t) \vec {\hat q}-C\vec {\hat f}\ee^{\ii\omega_f t}\rangle_{\overline \Omega}
+\sigma(\langle \vec {\hat f}, \vec {\hat f} \rangle_\Omega -1),
\label{eq:L}
\end{align}
where $\vec {\hat q}^+$ is the Lagrange multiplier in the form of adjoint perturbation fields to impose state constraints, and $\sigma$ is the Lagrange multiplier to constrain the force to unity magnitude. In the Lagrangian~(\ref{eq:L}), we have employed an instant $T_i$ at which $\hat q(T_i):=\hat q_0$ to remove $\hat q_0$ from the derivation and simplify the process. The first order optimality conditions for Lagrangian $\mathcal L$ dictates that variation of $\mathcal L$ with respect to forward, adjoint and control variables vanish identically (e.g. \cite{gunzburger03}), i.e.,
\begin{equation}
\frac{\partial \mathcal L}{\partial {\vec {\hat q}}}\delta\vec {\hat q}+
\frac{\partial \mathcal L}{\partial {\vec {\hat q^+}}}\delta\vec {\hat q^+}+
\frac{\partial \mathcal L}{\partial {\vec {\hat f}}}\delta\vec {\hat f}+
\frac{\partial \mathcal L}{\partial \sigma}\delta\sigma
=0,
\label{eq:optimality}
\end{equation}
where the directional variation is defined as, e.g., for the arbitrary variation $\delta\vec{\hat q}$ in state space,
\begin{equation}
\frac{\partial \mathcal L}{\partial {\vec {\hat q}}}\delta\vec {\hat q}=\lim\limits_{\epsilon \rightarrow 0}\frac{\mathcal L(\vec {\hat q}+\epsilon \delta \vec {\hat q} ,\vec {\hat q}^+,\vec {\hat f},\sigma)-\mathcal L(\vec {\hat q},\vec {\hat q}^+,\vec {\hat f},\sigma)}{\epsilon}.
\end{equation}
Setting the variations $\delta \vec {\hat q}^{+}=\delta \vec {\hat f}=0$, and letting $\vec {\hat q}'$ vary freely yields $\mathcal L_{\vec {\hat q}}(\vec {\hat q}')=0$. These equations are manipulated by utilizing integration by parts in space and time as many times as necessary until all differential operators on state fields are moved on to adjoint fields. The resulting boundary integrals in this process are eliminated by utilizing the homogeneous boundary conditions of OSS equations.

Variation of the Lagrangian with respect to each component of the forcing vector should vanish as a result of optimality condition in (\ref{eq:optimality}). Enforcing this stationarity condition yields for the streamwise component
\begin{align*}
2\frac{\partial \mathcal L}{\partial \hat f_u}\delta\hat f_u &=2\sigma\intz\hat{f}_u^*\delta\hat{f}_u\mathrm dz+\intt\intz(\hat w^+)^*\ii \alpha\frac{\partial \delta \hat f_u}{\partial z}\ee^{\ii\omega_f t}\mathrm dz \mathrm dt\\
&-\intt\intz(\hat\eta^+)^*\ii \beta\delta\hat {f_u} \ee^{\ii\omega_f t}\mathrm dz \mathrm dt +c.c.\\
&=2\sigma\intz\hat{f}_u^*\delta\hat{f}_u\mathrm dz+\left .\intt \left( (\hat w^+)^*\ii \alpha \delta\hat f_u\right) \right |_0^{\infty} \ee^{\ii\omega_f t}\mathrm dt\\
&- \intt\intz \frac{\partial (\hat w^+)^*}{\partial z} \ii\alpha \delta\hat f_u\mathrm dz \mathrm dt-\intt\intz(\hat\eta^+)^*\ii \beta\delta\hat {f_u} \ee^{\ii\omega_f t}\mathrm dz \mathrm dt +c.c. \\
&=\intz \delta \hat {f_u}  \left(2\sigma  \hat f_u^*+\intt \left(-\ii \alpha \frac{\partial (\hat w^+)^*}{\partial z}- \ii \beta (\hat\eta^+)^* \right)\ee^{\ii\omega_f t}\mathrm dt\right) \mathrm dz+c.c.=0,
\end{align*}
where we have made use of the Green's identity and the homogeneous adjoint boundary conditions $\hat w^+(0)=\hat w^+(z\rightarrow \infty)=0$. As the variation $\delta\hat {f_u}$ is a free variable, the optimality condition holds only if
\begin{equation*}
2\sigma  \hat f_u^*+\intt \left(-\ii \alpha \frac{\partial (\hat w^+)^*}{\partial z}- \ii \beta (\hat\eta^+)^* \right)\ee^{\ii\omega_f t} \mathrm dt=0.
\end{equation*} 
Manipulating the complex conjugates, we obtain the following expression for the streamwise component of the optimal force 
\begin{equation}
\label{eq:con1}
\hat f_u^{opt}:=\hat f_u=-\frac{1}{2\sigma}\intt \left(\ii \alpha \frac{\partial \hat w^+}{\partial z} + \ii \beta\hat\eta^+\right) \ee^{-\ii\omega_f t}  \mathrm dt.
\end{equation} 
The spanwise component of the optimal force is derived in a similar way:
\begin{align*}
2\frac{\partial \mathcal L}{\partial \hat f_v}\delta\hat f_v  &=2\sigma\intz\hat{f}_v^*\delta\hat{f}_v \mathrm dz+\intt\intz (\hat w^+)^*\ii \beta \frac{\partial \delta\hat f_v}{\partial z}\ee^{\ii\omega_f t}\mathrm dz \mathrm dt \\
&+\intt\intz(\hat\eta^+)^*\ii \alpha\delta\hat {f_v} \ee^{\ii\omega_f t} \mathrm dz \mathrm dt +c.c.\\
&=2\sigma\intz\hat{f}_v\delta\hat{f}_v\mathrm dz+\left .\intt  (\hat w^+)^* \ii \beta\delta\hat f_v \right |_0^{z_{\max}}  \ee^{\ii\omega_f t} \mathrm dt\\
&-\intt\intz  \frac{\partial (\hat w^+)^*}{\partial z}\ii \beta\delta\hat f_v \ee^{\ii\omega_f t} \mathrm dz \mathrm dt+\intt\intz(\hat\eta^+)^*\ii \alpha\delta\hat {f_v} \ee^{\ii\omega_f t} \mathrm dz \mathrm dt  \\
&=\intz \delta\hat {f_v}  \left(2\sigma  \hat f_v^*+\intt \left(-\ii \beta \frac{\partial (\hat w^+)^*}{\partial z}+ \ii \alpha(\hat\eta^+)^* \right)\ee^{\ii\omega_f t} \mathrm dt\right) \mathrm dz=0,
\end{align*}
which yields for the optimal spanwise force
\begin{equation}
\label{eq:con2}
\hat f_v^{opt}=\frac{1}{2\sigma}\intt \left(-\ii \beta \frac{\partial \hat w^+}{\partial z} + \ii \alpha\hat\eta^+\right) \ee^{-\ii\omega_f t} \mathrm dt.
\end{equation} 
Furthermore, the vertical component is derived as follows
\begin{align*}
2\frac{\partial \mathcal L}{\partial \hat f_w}\delta\hat f_w&=2\sigma\intz\hat f_w^*\delta\hat f_w\mathrm dz+\intt \intz  (\hat w^+)^* k^2 \delta\hat f_w \mathrm dz\mathrm dt \\
&=\intz \delta\hat {f_w}  \left(2\sigma  \hat f_w^*+\intt  k^2 (\hat w^+)^*   \ee^{\ii\omega_f t} \mathrm dt\right) \mathrm dz=0.
\end{align*}
Consequently, we obtain
\begin{equation}
\label{eq:con3}
\hat f_w^{opt}=-\frac{1}{2\sigma}\intt k^2 \hat w^+  \ee^{-\ii\omega_f t}\mathrm dt.
\end{equation} 
Finally, the variation with respect to $\sigma$ at a stationary point is
\begin{equation*}
\frac{\partial \mathcal L}{\partial \sigma}\delta\sigma=\delta \sigma (\langle \vec {\hat f}, \vec {\hat f} \rangle_\Omega -1)=0,
\end{equation*}
which restores the constraint equation for the amplitude of the forcing
\begin{equation}
\label{eq:con4}
\langle \vec {\hat f}, \vec {\hat f} \rangle_\Omega=1
\end{equation}
Equations (\ref{eq:con1})--(\ref{eq:con4}) represent a closed system of equations for the three forcing components and $\sigma$.

\bibliographystyle{jfm}
\bibliography{allpapers.bib}

\end{document}